\documentclass{aa}

\usepackage{graphicx}
\usepackage{float}
\usepackage{txfonts}
\usepackage{multirow}
\let\checkmark\relax
\usepackage{dingbat}
\usepackage{amsmath}      
\usepackage{xcolor}
\usepackage{color}
\usepackage{placeins}
\usepackage{url}
\usepackage{booktabs}
\usepackage[colorlinks=true,citecolor=blue]{hyperref}

\usepackage{pifont}

\newcommand{\Msun}{M$_\odot$}

\newcommand{\ha}{H$\alpha$}

\newcommand{\geza}[1]{\textcolor{black}{#1}}

\newcommand{\orcid}[1]{\href{https://orcid.org/#1}{\includegraphics[width=10pt]{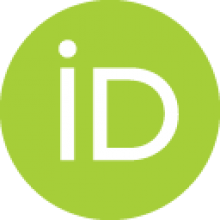}}}

\begin{document}

\title{Between Plateaus and Slopes: A Data-Driven Exploration of Spectral Diversity Across Type IIP/L Supernovae}
\titlerunning{Data-Driven view on SNe II spectral time series}

\author{ G. Cs\"ornyei, 
        \inst{1}\orcid{0000-0001-9210-9860}
        \and
        C. P. Guti\'errez
        \inst{2,3}\orcid{0000-0003-2375-2064}
        }

\institute{European Southern Observatory, Karl-Schwarzschild-Str. 2, 85741                Garching, Germany\\
        \email{Geza.Csoernyei@eso.org}  
        \and
        Institut d'Estudis Espacials de Catalunya (IEEC), 08860 Castelldefels (Barcelona), Spain 
        \and
        Institute of Space Sciences (ICE, CSIC), Campus UAB, Carrer de Can Magrans, s/n, E-08193 Barcelona, Spain. 
}

\date{}

\abstract
{Type II supernovae (SNe~II) have been traditionally separated into several subgroups based on their photometric and spectroscopic properties, but whether these represent distinct progenitors or a continuous distribution remains debated. Over the past decade, growing observational evidence has suggested a possible continuity between slow- (IIP) and fast-declining (IIL) SNe.}
{We investigate the continuity of the SNe~IIP/L subclasses through a data-driven statistical analysis applied to spectral time series, aiming to determine whether significant correlations exist between the overall spectral shapes and light curve decline rates.}
{We introduce a novel standardization method for SN II spectra. After empirically flattening the spectra via continuum normalization, we interpolate the resulting "feature spectra" onto a fixed grid of epochs using Gaussian Process regression. The interpolated spectra are then analyzed using Principal Component Analysis to explore correlations.}
{We find that SNe IIP and IIL form a continuum spectroscopically, though some clustering remains. The spectral diversity is characterized mainly by two components: one continuous group with well-defined P-Cygni profiles and another with "less-regular" features likely driven by enhanced circumstellar material (CSM) interaction. Our results reveal that the spectral diversity of SNe IIP/L diminishes over time. Comparisons with radiative transfer models confirm that both CSM interaction and hydrogen envelope mass variations are required to explain the diversity. We confirm observational correlations, namely that steeper light curve declines correspond to weaker spectral features, indicating that SNe~IIL tend to show weaker emission and, in some cases, a lack of distinct absorption lines. This trends break down by enhanced CSM interaction that modifies the P-Cygni profiles.}
{Our data-driven method reveals underlying spectral correlations and supports a continuous distribution between IIP and IIL subtypes, with the CSM interaction being one of the main drivers. This method paves the way for more refined classification algorithms.}

\titlerunning{A data-driven view on spectral diversity of SNe~II}
\maketitle

\section{Introduction}

Core-collapse supernovae (SNe) result from the catastrophic explosions of massive stars ($>8$ \Msun). They exhibit a broad range of spectral and photometric properties,  which serve as the basis for their classification into different spectral types. The most common type of core-collapse SNe is Type II supernovae (SNe~II; \citealt{Li11, Shivvers17}), which are characterized by optical spectra exhibiting strong Balmer lines \citep{Minkowski41}. These events originate from massive progenitor stars that retain a significant amount of their hydrogen-rich outer envelopes at the time of the explosion. 

For decades, SNe~II were divided into two main subtypes based on their light-curve morphology: SNe~IIL and SNe~IIP \citep{Barbon79}. SNe IIL exhibit a rapid decline in brightness after peak, while SNe IIP display a plateau phase, characterized by a nearly constant luminosity for several weeks. Beyond these, three additional subclasses of SNe~II have been identified, distinguished either by their light curve behavior or spectral features. The so-called 1987A-like events display slowly rising light curves, reminiscent of the prototypical SN 1987A \citep{Taddia12, Taddia16}. SNe~IIb are transitional objects that evolve from hydrogen-rich (SNe~II) to hydrogen-poor (SNe~Ib) spectra \citep{Filippenko93}. In contrast, SNe IIn, which are the brightest subtype of SNe II, exhibit prominent, long-lasting narrow hydrogen emission lines, attributed to strong interaction \citep{Schlegel90}. 

A long-standing question in the SN field has been whether these subclasses are distinct due to separate evolutionary paths or whether they represent a continuous distribution of events, which appear distinct only because of the wide parameter range and the relatively low number of events compared to it. Recent studies based on larger and more homogeneous datasets have revealed continuous distributions in photometric properties, supporting the idea that SNe~IIL and IIP likely originate from a common progenitor population \citep{Anderson2014a, Sanders15, Galbany16, Valenti16, Gutierrez2017}. The most favored explanations for such continuity are the variability in the mass of the hydrogen envelope of red supergiant progenitors \citep[e.g.][]{Eldridge2018, Hillier2019, 2025arXiv250714665F} or the extent of circumstellar material (CSM) interaction ongoing during the explosion \citep[e.g.][]{Hillier2019, Dessart2024}. Alternatively, the phase of progenitor pulsations at explosion epoch was also proposed as a potential driver for the diversity \citep{Bronner2025}.

CSM interaction has become a central focus in SN~II research, especially following the discovery of flash ionization features resembling the spectral signatures of SNe~IIn \citep{Yaron2017, Bruch2021, 2025arXiv250504698J, Dessart2025}. Studies suggest that even regular SNe~IIP/L also exhibit signs of at least a mild CSM interaction \citep{Morozova2017, Dessart2022}. More recently, \cite{Hinds2025} and Ertini et al. in prep. demonstrated that early light curves of SNe~IIP, IIL and IIn form a continuous distribution in both rise times and inferred CSM properties. This is also supported by the findings of \cite{Bruch2023}, who found that SNe~II showing more persistent flash ionization features (lasting for weeks or more) tend to be systematically brighter, potentially bridging the gap between typical SNe~ IIP/L and strongly interacting SNe~IIn. These CSM-oriented investigations can be further extended towards late-time observations by modeling dust properties of SNe~II, which have also been proposed to be connected to CSM interaction \citep{2025arXiv250722763T}.

Similar studies have been conducted to understand the continuity in the context of hydrogen envelope mass, particularly regarding SNe~IIb. While population synthesis models predict a continuous origin \citep{Eldridge2018}, observational evidence has been mixed. Based on light-curve analyses, \cite{Pessi19} found no compelling evidence supporting this continuity \citep[also see,][]{Arcavi12}. In contrast, \cite{Gonzalez-Banuelos25} revealed that the spectral properties of SNe~II and IIb do exhibit a continuous distribution. This interpretation aligns with the models of \cite{Dessart2024-2}, which demonstrated that a continuous population of Ib, IIb, and IIP/L SNe can arise from binary progenitor systems with varying initial separations. Further supporting this picture, \cite{Ercolino2025} showed that mass transfer in binaries can explain the CSM interaction signatures observed in stripped envelope SNe.

These findings highlight the importance of investigating spectral and photometric continuity across SN~II subclasses. With the advent of future and ongoing wide-field surveys, such as the Zwicky Transient Facility \citep[ZTF;][]{Bellm2019, Graham19} and the forthcoming Vera Rubin Observatory Legacy Survey of Space and Time \citep[LSST;][]{LSST}, complemented by extensive follow-up programs, the number of observed SNe~II is rapidly increasing. This presents a unique opportunity to identify and fill gaps between SN II subclasses.

To fully exploit these datasets, robust data-driven methods are needed. This approach, however, faces a difficulty: the spectra of SNe~II are intrinsically complex, containing multiple features across the optical wavelengths and evolving substantially over time. These characteristics make it challenging to apply techniques used for galaxy or stellar spectra without dealing with the evolution of the spectra. Furthermore, the sampling of the spectra for the individual objects also introduces a limitation, as not all SNe are observed at the same epoch or with the same cadence. Overcoming these issues requires methods that facilitate epoch-wise comparisons of spectra directly, as done, for example, in \cite{Lu2023} or \cite{Burrow24}.

In this work, we investigate the spectral diversity and continuity within the SN~II population more closely using data-driven techniques. We perform this by applying interpolation and dimensionality reduction techniques, namely Gaussian Process-based time series reconstruction and Principal Component Analysis, to uncover underlying patterns and correlations. Given the complexity of SN~II spectra among the different subtypes, we limit our analysis to the more classical SNe~IIP/L population. Throughout this manuscript, the term “SNe II” refers specifically to the IIP/IIL population. With this in mind, our primary goal is to investigate whether a strong correlation exists between the spectral appearance of SNe IIP/L and the decline rate of their light curves.

The paper is organized as follows. A description of the data is presented in Section~\ref{sec:data}. The methodology is given in Section~\ref{sec:process}. In Section~\ref{sec:analysis}, we present the analysis. Finally, in Section~\ref{sec:disc} and Section~\ref{sec:conc}, we present the discussion and conclusions, respectively.

\section{Data}
\label{sec:data}

To ensure the reliability of the spectral interpolation and to minimize potential biases, we selected SNe with a sufficient number of spectra during the early photospheric phase (up to 55 days after explosion). Specifically, we required a minimum of three spectra per object, with no more than 10 days between consecutive observations, to reduce interpolation artifacts (see Section~\ref{sec:analysis}). The final sample was compiled from various sources, most notably the Carnegie Supernova Project (CSP-I; \citealt{Gutierrez2017a, Gutierrez2017}) and the CfA Supernova Program \cite{Hicken17}. Additionally, we included several well-observed objects from individual studies, all of which are detailed in Table~\ref{tab:sample1}. 

In addition to requiring a minimum number of spectra with good cadence, we applied quality criteria to the individual spectral observations. Spectra were excluded if they covered only a narrow wavelength range or if they were excessively dominated by noise in comparison to other epochs for the same object. The final sample consists of 147 SNe~II, comprising a total of 1355 spectra, with a median of 6 spectra per object. Moreover, we restricted our sample to objects for which the explosion epoch could be established with an uncertainty of a maximum of 3-5 days (although the majority of our sample, especially the additions from the last decade, have substantially more precise estimates than this limit). In the final sample, 86 objects were taken from CSP-I, 18 from CfA (eight objects were covered by both), and the remaining 51 were taken from the literature. When available, the explosion epochs were taken from the literature \cite[e.g.][]{Gutierrez2017a}, typically defined as the midpoint between the last non-detection and the discovery. In cases where sufficient photometric coverage existed, we independently estimated the explosion epoch by fitting the early light curve as described in \cite{Csoernyei2023}. Explosion epochs are also presented in Table~\ref{tab:sample1}.

\section{Pre-processing spectral time series}
\label{sec:process}

To enable a meaningful comparison across different supernovae, we applied a series of pre-processing steps to minimize systematic differences between spectra. First, we restricted all spectra to the wavelength range of 3500 -- 9500 $\AA$ and masked narrow host-galaxy emission lines from the Balmer series, [OII], [OIII], [NII], and [SII], as well as telluric lines when present.

Beyond these, multiple factors still complicate statistical analysis of SN spectra, such as the reddening, the imperfect flux calibration, fringing effects caused by internal reflections in the optical system, and random noise. While the first two, reddening and flux calibration issues, can be removed or limited through using Galactic dust maps and recalibrating the spectra to match the photometry (e.g. \citealt{Csoernyei2023}), both introduce their own uncertainties (e.g., dust maps accuracy, host-galaxy extinction, and photometric system differences and inaccuracies). Since both reddening and flux calibration primarily affect the continuum in a multiplicative manner, we removed these effects empirically through continuum normalization, in a similar fashion as done in \cite{Blondin2007}. This yields a "residual spectrum" primarily containing intrinsic absorption and emission features (see Section~\ref{sec:contrem}).

Following the continuum removal, we further reduced supernova-to-supernova variation by mitigating residual noise and fringing patterns. Although dimensionality reduction techniques can help distinguish spectral features from noise, structured artifacts like fringing, if present across multiple spectra, can bias the analysis. To address this, we applied Gaussian Process interpolation to denoise the spectra. This method was specifically adapted to the characteristics of our sample, and the corresponding procedures are described in Section~\ref{sec:denoise}.

\subsection{Continuum removal}
\label{sec:contrem}

To remove the continuum component from the observed spectra, we employed \textit{locally weighted scatterplot smoothing} (LOWESS), a technique functionally similar to the Savitzky-Golay filter. LOWESS generalizes moving average and polynomial regression, enabling non-parametric fitting that can act as a low-pass filter for the spectra, following only those spectra features that encompass larger wavelength ranges. While it is not as flexible as, for example, Gaussian Processes for separating signal from noise, it offers a robust and computationally efficient method for continuum estimation.

\begin{figure*}
\centering
\includegraphics[width=0.8\linewidth]{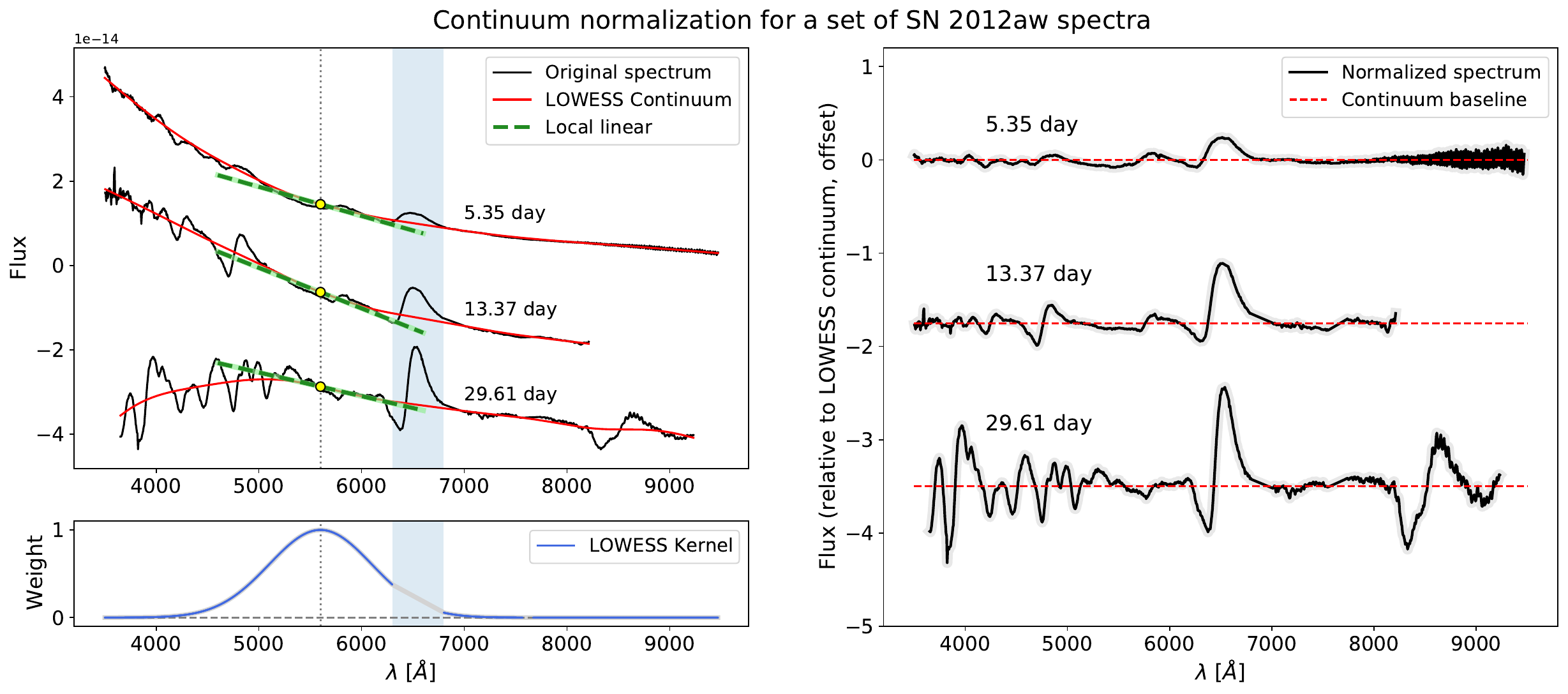}
\caption{Example of the LOWESS-based continuum removal applied to selected spectra of SN~2012aw. \textbf{Top left:} Original spectra with the LOWESS-smoothed continuum overplotted. The yellow marker indicates a reference wavelength at which the local linear fits are displayed for each spectrum. The blue shaded region highlights the masked \ha\ feature, excluded from the continuum fitting to avoid bias \textbf{Bottom left:} Shape of the LOWESS kernel at the reference wavelength, illustrating the weighting scheme used during smoothing. \textbf{Right:} Continuum-normalized spectra obtained by dividing the original data by the fitted LOWESS continuum.}
\label{fig:continuum}
\end{figure*}

For our analysis, we applied a first-degree polynomial implementation of the filter, which is well-suited to track the smooth, gradual variations of the continuum. Fig.~\ref{fig:continuum} (left panel) shows examples of this filtering applied to spectra at various epochs. Furthermore, to ensure that only large-scale trends are captured, we set the smoothing kernel width to 500 $\AA$ngstr\"oms (which corresponds to the $1\sigma$ width for the Gaussian). However, because the LOWESS filter is based on local low-order polynomial fits, which are sensitive to the local flux average, broad spectral features that move away significantly from the continuum level can introduce biases into the fit. One notable example is \ha, which often expands hundreds of $\AA$ngstr\"oms and typically exhibits the emission significantly outweighing the absorption line. To prevent this from skewing the continuum estimation, we masked the \ha\ region during the fitting process, as shown in Fig.~\ref{fig:continuum}.

\subsection{Denoising of spectra}
\label{sec:denoise}

After normalizing each spectrum by its continuum, we performed a denoising step to minimize the influence of noise on subsequent analyses. Although Principal Component Analysis (PCA, \citealt{Pearson1901}) is generally effective at isolating random noise, we opted to denoise beforehand due to specific characteristics of our sample: many spectra,  particularly at earlier phases, exhibit higher noise levels in the red end due to relatively lower fluxes, and spectroscopic fringing signatures are also present in several cases (e.g. the spectrum of SN~2012aw at 5.35 days shown in right panel of Fig.~\ref{fig:continuum}). These non-random patterns can contaminate the principal components if not addressed in advance.

To mitigate this contamination, we fitted the normalized spectra with a combination of Gaussian Processes\footnote{Gaussian Processes or GP are a nonparametric supervised learning method highly useful for regression applied on irregular data. For more details on the technique, we refer the reader to \url{https://scikit-learn.org/stable/modules/gaussian_process.html}.} (GP; \citealt{Rasmussen06}),  using the \textsc{george} Python package. To capture the variability of the spectra, we split each spectrum at 7000 $\AA$ into a blue and a red region, allowing for different GP kernel configurations tailored to the noise properties on each side. This division was motivated by the fact that fringing affects broader wavelength scales, particularly in the red, while random noise affects the blue region more at later phases, which contains finer spectral features (e.g. line blanketing features), and that would be lost with overly broad smoothing. For each segment, we modeled the data using two additive GP components, one for the signal and one for the noise:

\begin{itemize}
\item{\textbf{Noise model:} For the noise, a product of a sinusoidal kernel (to capture oscillatory noise) and an exponential kernel (to model variations in noise amplitude, e.g., those introduced by the continuum normalization).}

\item{\textbf{Signal model:} Two Matern-32 kernels with two different length scales: one short (to capture features such as emission and absorption lines), and a long one (to follow smoother variations such as residual continuum structure).}
\end{itemize}

\begin{table}
\tiny
\centering
\setlength{\tabcolsep}{3pt}
\begin{tabular}{|c|c|c|c|c|c|}
\hline
\textbf{Term}      & \textbf{Kernel} & \multicolumn{2}{|c|}{\textbf{Blue side}}  & \multicolumn{2}{|c|}{\textbf{Red side}} \\
\hline
                   & ExpSine2        & $\gamma = 100$    & $\log{P} = 0.1$       & $\gamma = 100$    & $\log{P} = 0.1$     \\
\textbf{Noise}     &     *           &                   &                       &                   &                     \\
                   & ExpSquared      & $c = 100$         & $l = 2000$            & $c = 10$          & $l = 500$           \\
\hline
\hline
                   & Matern32        & $c = 100$         & $l = 10^5$            & $c = 100$         & $l = 10^5$          \\
\textbf{Continuum} &     +           &                   &                       &                   &                     \\
                   & Matern32        & $c = 20$          & $l = 10^4$            & $c = 20$          & $l = 5000$          \\
\hline
\end{tabular}
\vspace{0.2cm}
\caption{Parameters used for Gaussian Process denoising. The listed values correspond to the kernel input parameters defined by the \textsc{george} Python package. Specifically: $\gamma$ represents the correlation/randomness parameter of the sinusoidal kernel; $\log{P}$ denotes the wavelength scale of the variation component; $c$ is a scaling factor that multiplies the kernel amplitude; $l$ is the characteristic length scale of the kernel, controlling the smoothness of the modeled function. For detailed definitions and kernel formulations, we refer the reader to the official documentation of the \textsc{george} package\protect\footnotemark.}
\label{tab:parameters}
\end{table}

\begin{figure}
\centering
\includegraphics[width=0.9\linewidth]{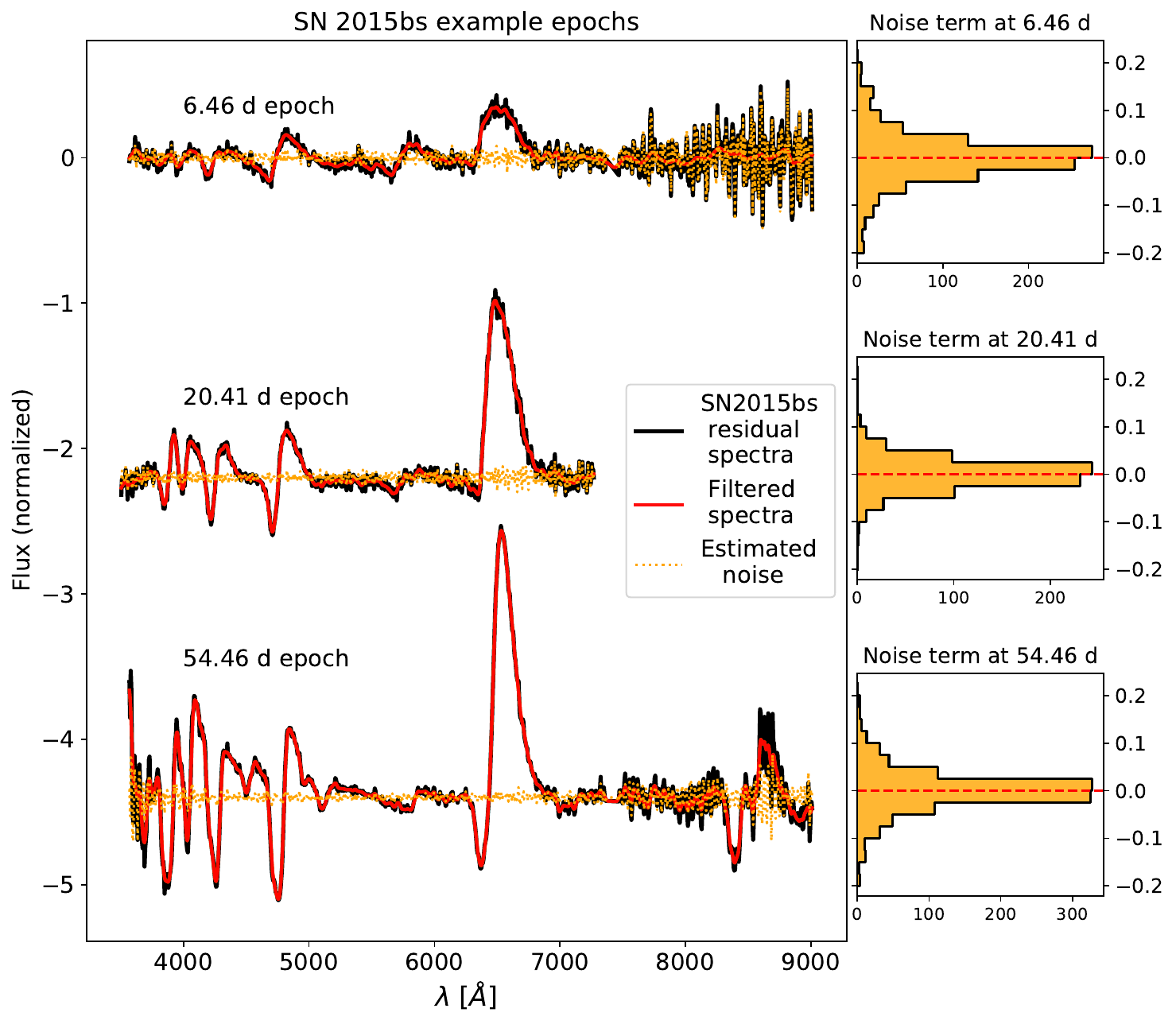}
\caption{Gaussian Process denoising applied on SN~2015bs example spectra. The left plot shows the original spectra overlaid with their denoised counterparts, demonstrating the effectiveness of the procedure in preserving spectral features. The right panel shows histograms of the flux residuals attributed to noise. The near-Gaussian distribution of these residuals supports the assumptions of white noise and confirms that no significant correlated features were removed during the process.}
\label{fig:example_denoising}
\end{figure}

The specific kernel parameters used for the denoising are listed in Table~\ref{tab:parameters}. Each spectrum was fit on the blue and red sides using the sum of the corresponding GP kernel combinations, effectively separating noise from the underlying spectral features. Fig.~\ref{fig:example_denoising} shows a representative outcome of the denoising process. To verify that only noise was removed, we examined the distribution of the extracted noise term. Assuming white noise, this distribution should follow a Gaussian profile; any significant deviation would suggest that parts of the actual spectrum were accidentally modeled as noise. As shown in the figure, the noise component closely matches a Gaussian distribution, indicating that the denoising procedure effectively suppresses noise without significant losses in the spectrum. This denoising step was applied to all continuum-normalized spectra. All subsequent analyses were carried out on the denoised data. For reference, we also performed the full analysis on the original (non-denoised) dataset, and the test results are presented in Appendix \ref{app:noisyrun}.

\addtocounter{footnote}{-1}
\stepcounter{footnote}\footnotetext{\url{https://george.readthedocs.io/en/latest/user/kernels/}}

\section{Analysis}
\label{sec:analysis}

\subsection{Standardizing the spectra to given epochs}

\begin{figure}
\centering
\includegraphics[width = 0.9\linewidth]{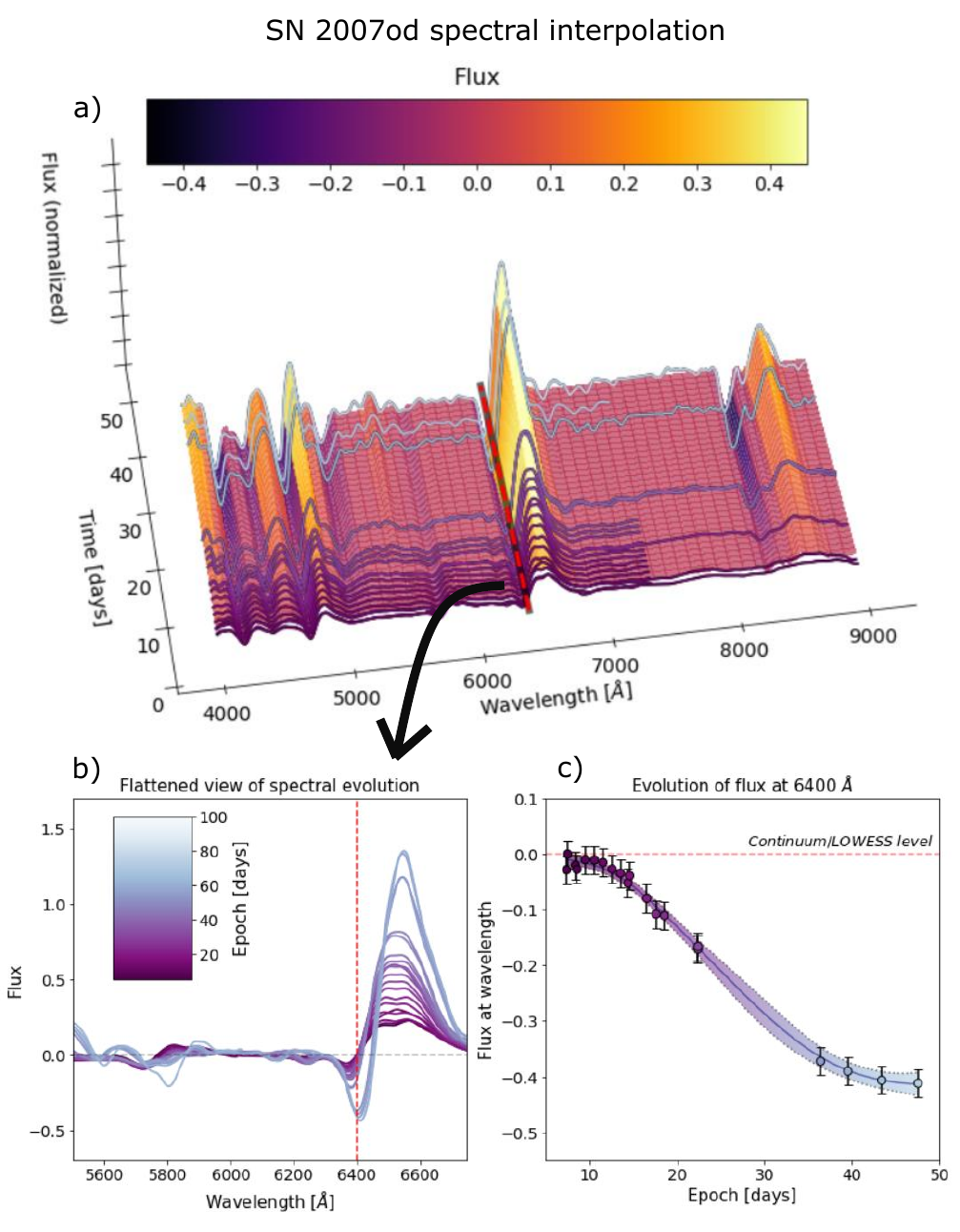}
\caption{Illustration of the Gaussian Process-based interpolation of the spectral time series. \textbf{Panel (a)} 3D view of the continuum-normalized spectral time series of SN~2007od interpolated onto a common time–wavelength grid. \textbf{Panel (b)} projection of panel (a), collapsed along the time axis, highlighting the evolution of spectral changes in the region around \ha. \textbf{Panel (c)} flux evolution at a given wavelength bin (red dashed line in panels (a) and (b)), with GP fit color-coded by epoch.}
\label{fig:interpolation}
\end{figure}

This study investigates spectral diversity and continuity across the SN IIP/L subtypes. Previous analyses of SNe~II have suggested that these events form a continuous class, with no clear dichotomy in either their light curve or spectral properties (see e.g. \citealt{Anderson2014a, Anderson2014b, Gutierrez2014, Gutierrez2017,Gutierrez2017a}). While spectroscopic differences have been observed, these variations appear continuous rather than two separate subclasses. For instance, \cite{Gutierrez2014, Gutierrez2017} showed a correlation between the shape of the \ha\ P-Cygni profile and the light curve decline rates, with more rapidly declining SNe~II showing suppressed \ha\ absorptions. This trend is also evident in our sample.

One challenge in identifying robust spectral trends related to the light curve decline rates lies in the variation of observational epochs across different SNe. To mitigate this, previous works such as \cite{Anderson2014b} and \cite{Gutierrez2017} focused on single spectral features (e.g., peak blueshifts or \ha\ absorption-to-emission ratios), which were then interpolated to a common epochs for comparison. In this work, rather than limiting the analysis to a single spectral parameter, we undertake a full spectral time-series comparison. To do this, instead of interpolating a \emph{single parameter}, we interpolate the \emph{entire spectral time evolution} across a range of common epochs. Our method follows the GP interpolation approach similar to that outlined in \cite{Saunders2018} and \cite{Leget2020}. First, each spectrum was rebinned onto a uniform wavelength grid spanning 3500 and 9500 $\AA$. We then applied GP independently to each wavelength bin in the time series of each SN. As in \cite{Saunders2018}, we did not enforce correlations between neighboring wavelength bins, both to reduce computational cost and because including such correlations did not significantly improve interpolation accuracy, particularly given the intrinsic spectral diversity of SNe~II. Fig.~\ref{fig:interpolation} illustrates the GP-based spectral interpolation process. Using this method, we reconstructed complete spectral sequences across the early photospheric phase and interpolated spectra for each SN at four common epochs: 20, 30, 40 and 50 days post-explosion. These interpolated spectra are then used for the subsequent comparative analyses.

\subsection{Dimension reduction via PCA}

\begin{figure}
\centering
\includegraphics[width = 0.95\linewidth]{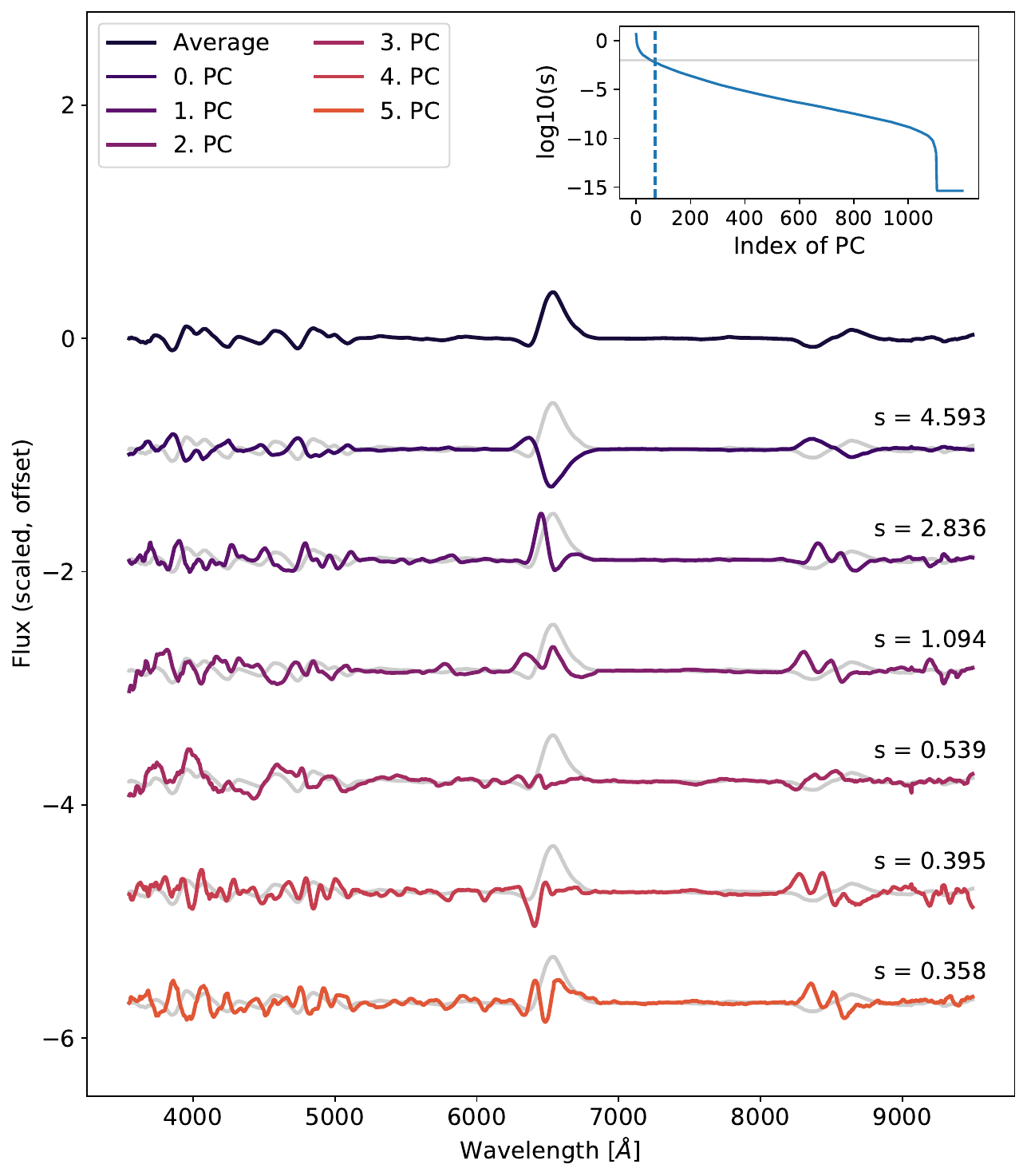}
\caption{PCA results at one representative epoch. The figure shows the PCA applied to continuum-normalized spectra at 20 days post-explosion. The top spectrum represents the mean SN~II spectrum at that epoch, while the spectra below show the first few eigenspectra derived from the PCS (via singular value decomposition, SVD), ordered from top to bottom by decreasing significance. For reference, the mean spectrum is overplotted in light gray behind each eigenspectrum. The numbers next to each eigenspectrum indicate the corresponding eigenvalue, which reflects the variance explained by that PC. The inset plot in each panel displays the eigenvalue decay, indicating the variance captured by each component.}
\label{fig:PCA_eigenvectors}
\end{figure}

To reduce the dimensionality of the spectral dataset and facilitate the comparison between objects, we applied PCA to the spectra at each selected epoch independently. PCA is widely used in astronomy for dimensionality reduction and identifying the directions of greatest variability within a dataset \citep[see e.g.,][]{Hsiao07, Muller-Bravo22, Burrow24, Aamer25, Muller-Bravo25}. Although PCA can be suboptimal for spectra where key features shift significantly in wavelength (such as SN features, where the photospheric velocity introduces a range of Doppler shifts), we used it here due to its ease of implementation and interpretability. Fig.~\ref{fig:PCA_eigenvectors} shows the results of the PCA applied to the sample spectra at 20 days post-explosion (Figure~\ref{fig:PCA_eigenvectors40d} shows the 40-day epoch case). At all epochs considered (20, 30, 40 and 50 days post-explosion), the derived eigenspectra reveal that spectral variability spans multiple wavelength regions: none of the most significant components isolate changes in a single feature alone.
This implies strong correlations between spectral features at each epoch, suggesting a coherent evolution across the spectrum.

Another noteworthy trend is observed in the distribution of the singular values (Table~\ref{tab:PCA}), representing the proportion of variance explained by each principal component. Although direct comparison of singular values across PCA runs is limited (due to differences in the number of input spectra at earlier epochs), we observe a general increase in total variance towards later epochs. This is consistent with the emergence and strengthening of spectral lines over time. Additionally, at later epochs, the variance becomes more concentrated in the first few components, indicating that the spectral diversity decreases with time, and a smaller number of components can capture the bulk of the variance. 
As we discuss later, one possible explanation for the observed excess variability is the early-time interaction between the ejecta and CSM \citep{Gutierrez2017a}, after which a "normal" SN II spectrum can emerge \citep[e.g.][]{Hillier2019}. Still, differences in the SN initial conditions, such as explosion energy or density profile, might also yield more pronounced differences at earlier phases \citep{Valenti16}.

From the PCA decomposition, we obtained principal component coefficients for each SNe at each epoch. These coefficients serve as a compressed, low-dimensional representation of the original spectra. For the subsequent analysis, we saved the first 70 coefficients per epoch, which were found to account for $\sim 99\%$ of the total variance. The remaining higher-order components primarily reflect random noise or negligible features.

\subsection{Clustering and groups within SN~II space}
\label{sec:clustering}

The PCA decomposition provides a compact, lower-dimensional representation of the spectral diversity in our sample, enabling the identification of potential clustering. Our primary objective is to determine whether slow- and fast-declining SNe exhibit a distinct break in their spectral appearance, resulting in clear dichotomies in the PCA coefficients, or whether they form a continuous distribution in the spectral parameter space. Given the high dimensionality of the PCA coefficients (70 per object per epoch), a two-dimensional projection is necessary for visual interpretation. Several nonlinear dimensionality reduction algorithms exist for this purpose, such as t-distributed Stochastic Neighbor Embedding (t-SNE; \citealt{Van08}). t-SNE an unsupervised algorithm that is highly effective at preserving local relationships between data points. At the same time, it enhances distinctions in high-dimensional space as well. Unlike PCA, t-SNE is a nonlinear method that does not yield easily interpretable parameters or axes; however, it offers a powerful tool for revealing latent structures and clusters.

\begin{figure*}
\centering
\includegraphics[width = 0.82\linewidth]{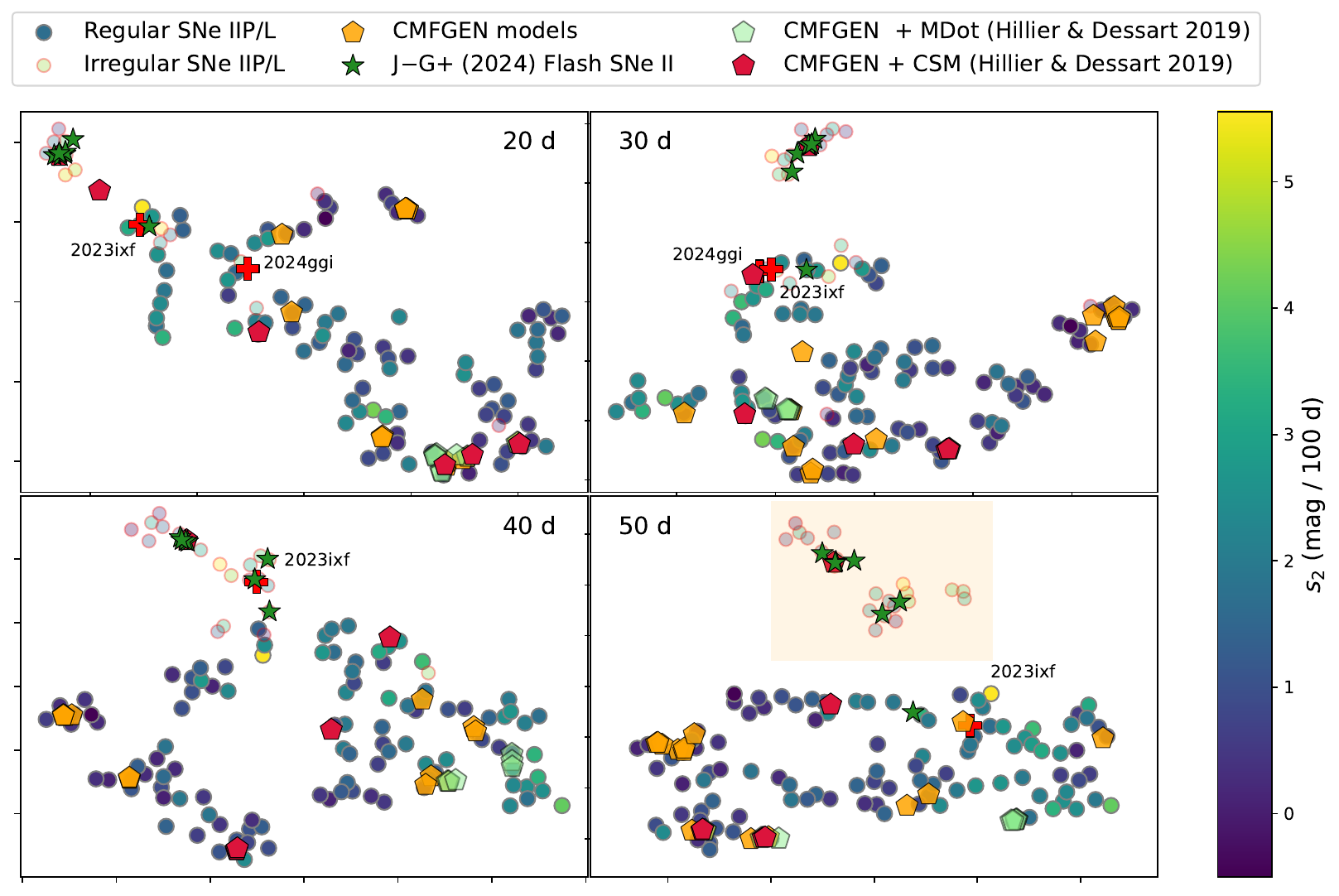}
\caption{t-SNE projection plots of the PCA coefficients (per epoch). These plots show how the SNe are distributed within the high-dimensional PCA space after dimensional reduction. Colored pentagons mark the positions of different \textsc{cmfgen} models variants. SNe~2023ixf and 2024ggi are also added using the spectra currently available in the literature (red crosses), along with CSM-interacting SNe~II from \cite{Jacobson-Galan2024} (green stars). A clear separation by $s_2$ values is observed across all epochs, confirming the link between spectral and light curve properties. At 50 days, two groups emerge: a continuous "regular SNe IIP/L" sequence along $s_2$, and a scattered subgroup, likely shaped by strong CSM interaction (marked by the yellow rectangle). Model projections align with expectations: CSM-rich models explain the subgroup objects, while mass-loss models resemble fast-declining SNe.}
\label{fig:tsne}
\end{figure*}

We implemented t-SNE using the \textsc{opentsne} package \citep{Policar2024}, which offers improved computational efficiency and the ability to embed new data points into existing t-SNE regions. We applied this technique to the PCA coefficients set at 20, 30, 40, and 50 days post-explosion. The result projections are presented in Fig.~\ref{fig:tsne}, using consistent hyperparameter settings across epochs. As shown in the figure, the progression of clustering with time is evident. At day 20, most SNe form a single, diffuse cloud with no well-pronounced outliers. Between days 30 and 40, this cloud begins to fragment, and by day 50 a clearer separation emerges, revealing a subgroup of SNe with non-standard spectra (highlighted by the yellow box in the bottom right panel of Fig.~\ref{fig:tsne}), and a continuous group of "spectroscopically normal" (canonical) SNe~IIP/L. At days 30 and 40, it is possible to recognize a subgroup of objects with weak (suppressed) spectral features and fast-declining light curves (corresponding to canonical IIL-s, middle of top right panel of Fig.~\ref{fig:tsne}) and a small subset of slow-declining, low-luminosity SNe~II (right center of top right panel of Fig.~\ref{fig:tsne}). By day 50, these two subsets merge into the "continuous" group, indicating that their spectral differences are less pronounced than those characterizing the marked subgroup.

We tested whether the emergence of this subgroup could be attributed to random alignment or to \geza{a robust feature} of our sample by applying both trimming and bootstrapping. In the trimming approach, we randomly removed 5\% (10\%) of the objects from the sample before applying t-SNE. For the bootstrapping test, we have drawn 63\% of the full dataset, allowing for duplicates. Clusters were identified using Gaussian Mixture Models (GMM) from the \texttt{sklearn} package. We recovered comparable groups containing at least 50\% of the original subgroup members and less than 25\% contamination in 92\% (74\%) of the trimming realizations. A subgroup matching the conditions also emerges in $\sim$ 60\% of the bootstrapping cases. 
\geza{These resampling tests indicated that the identified subgroup is not driven by a small number of individual objects, but instead shows moderate stability under perturbations of the dataset. This is consistent with a weakly separated low-count structure in the present sample. We emphasize, however, that the bootstrap recovery fraction derived above should be interpreted as a measure of robustness under resampling, rather than as a formal assessment of statistical significance.}\footnote{\geza{It has been shown that clustering stability depends strongly on the properties of the data and on the degree of separation between structures. In particular, resampling procedures can reduce the recovery rate even for meaningful clusters, particularly when they are small or only weakly separated from the bulk of the sample (e.g., \citealt{Ben-Hur2002}; \citealt{Hennig2007}; \citealt{vonLuxburg2010}). Hence, the recovery rates of low-population structures may not reach 100\% under resampling, as such clusters are disproportionately affected \citep[e.g.,][]{Hennig2007}.}} We explore the interpretations of this subgroup in more detail in Sec.~\ref{sec:disc}.

To expand the comparison through t-SNE, we included observational data of SN~2023ixf and SN~2024ggi, as well as \textsc{cmfgen} SN~II models \citep{Dessart13, Dessart17, Lisakov17, Hillier2019}, along with CSM interacting SNe~II from \cite{Jacobson-Galan2024}, which exhibited strong flash features. \footnote{We included "gold" and "silver" sample objects, which had a sufficient amount of spectra for our GP spectral interpolation. In the end, the included objects were SNe~2017ahn, 2019ust, 2020abjq, 2020pni, 2021tyw, 2022ffg and 2022pgf.} To include these events, we estimated their PCA coefficients based on the main sample and then projected them onto our 2D t-SNE embedding using \textsc{opentsne}. For SN~2023ixf, we used data from \cite{Zheng25}, while the SN~2024ggi data were obtained from \cite{Ertini25}. Due to the lack of spectroscopic coverage between 30 and 50 days for SN~2024ggi, it is only included in the 20- and 30-day projections.

Both SN~2023ixf and SN~2024ggi are located at the interface between the main population (regular SNe) and the detached subgroup. This placement is consistent with signatures of early CSM interaction, which delays and suppresses the appearance of P-Cygni profiles. Given the strong suppression or lack of P-Cygni features at early stages, both objects are mixed with canonical IIL-s in Fig.~\ref{fig:tsne}. In the case of SN~2023ixf, the emergence of more typical features after day 30 results in its migration toward the main group, though the absorption and emission features remain weaker than average. A similar transition is expected for SN~2024ggi, although this cannot be directly confirmed due to the missing epoch coverage. 

The positions of the \textsc{cmfgen} models align well with the various sub-populations of SNe IIP/L and reproduce our expectations of the models. The majority of the models from \citep{Dessart13, Dessart17, Lisakov17} have projected positions on the t-SNE plots that align well with the more canonical IIP-type SNe, with only a number of models scattering among IIL-like SNe that exhibit steeper light curve decline rates. The situation changes with the addition of models from \cite{Hillier2019}, which include model runs with CSM interaction through density profile modifications and mass loss. The models that assume mass loss of the progenitor before the explosions (with model names as $"x$*$p$\emph{0}$"$ in \citealt{Hillier2019}) are placed among the SNe that exhibit relatively high decline rates of $s_2 \sim 3 $~mag~/~100d. Interestingly, these models are the most similar to one another, which could imply that the variations caused by mass loss and reduced hydrogen envelope cannot account for all the diversity observed in the IIL subclass (at least within the modeling framework established in \citealt{Hillier2019}).

\begin{figure}
\centering
\includegraphics[width=0.95\linewidth]{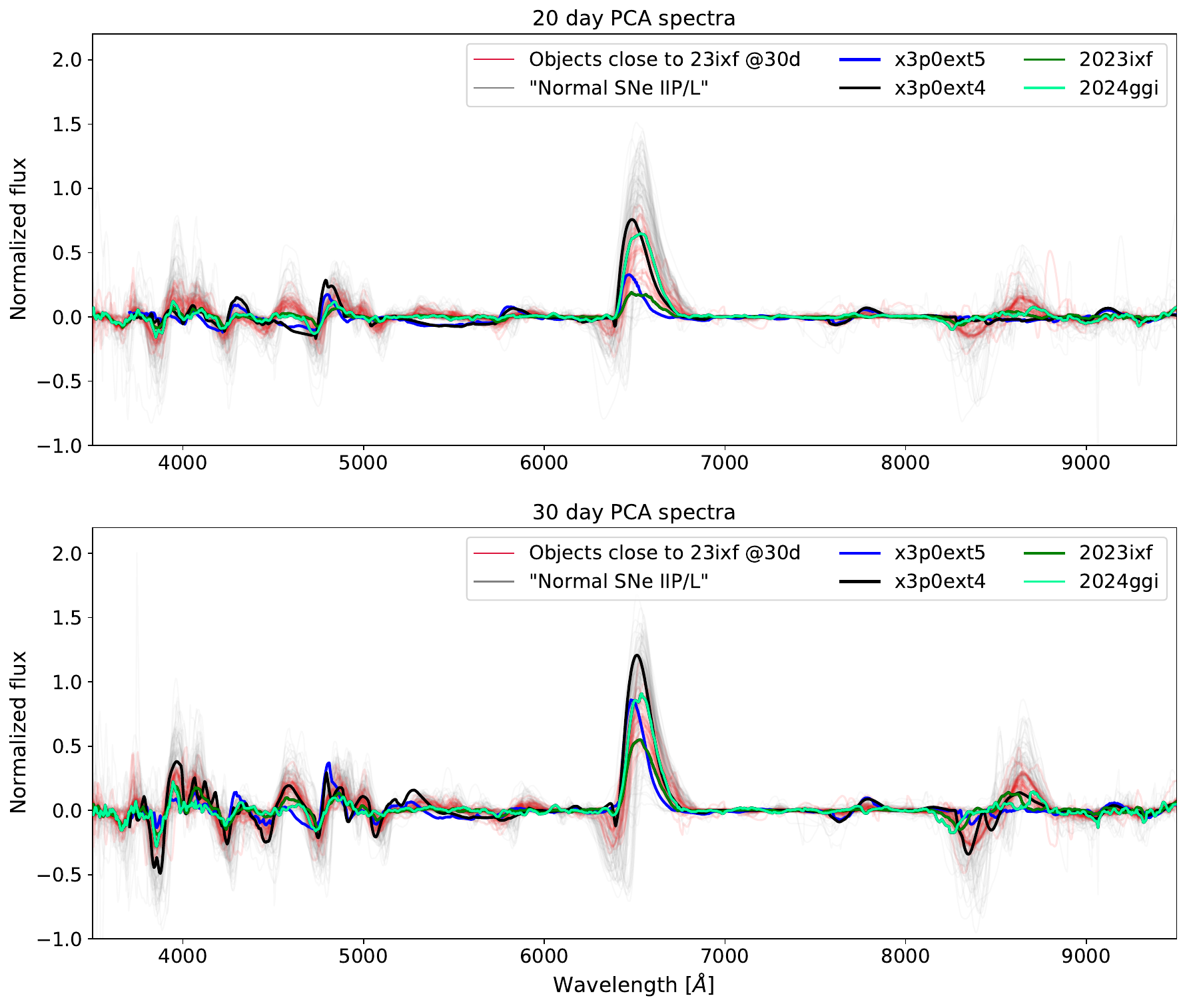}
\caption{Comparison between the day 20 and day 30 interpolated spectral features of 2023ixf and 2024ggi with the nearest models from \cite{Hillier2019}, based on  Fig.~\ref{fig:tsne}.}
\label{fig:23ixfand24ggi}
\end{figure}

Furthermore, these models exhibit notable differences in spectral appearance compared to SNe~2023ixf and 2024ggi. In contrast, models assuming CSM interaction can produce spectra that closely resemble those of these two SNe. Fig.~\ref{fig:23ixfand24ggi} shows the comparison of the spectra of SNe~2023ixf and 2024ggi at 20 and 30 day epochs with the models from \cite{Hillier2019} that fall the closest to them on the t-SNe plots corresponding to these epochs. The first best match is given by \texttt{x3p0ext5}, with \texttt{x3p0ext4} also yielding an adequate comparison to SN~2024ggi at 20 days. These models assume an intermediate amount of CSM interaction in the model grid. Given that the early-time behavior of the above two SNe is explained as being influenced by CSM interaction, the recovered match with the models that assume a similar behavior demonstrates the reliability of our projecting and clustering method.

Finally, the CSM interacting objects from \cite{Jacobson-Galan2024} are found to cluster almost exclusively within the detached subgroup, together with a single CMFGEN model (\texttt{x3p0ext6}), which assumes the highest scale of interaction with CSM. This overlap strongly supports the interpretation that the subgroup consists of SNe that experienced enhanced CSM interaction. Although the spectra within this group display considerable diversity, a common characteristic is the significant delay in the emergence of P-Cygni features in the spectra, which is usually attributed to CSM interaction. Furthermore, this comparison shows well that SNe exhibiting prevalent and strong flash features, hence having gone through enhanced early CSM interaction, will appear significantly different even at later phases, well after the CSM has been swept up by the ejecta.

\subsection{Measurement of reference parameters}
\label{sec:refpms}

\begin{figure*}
\centering
\includegraphics[width = 0.78\linewidth]{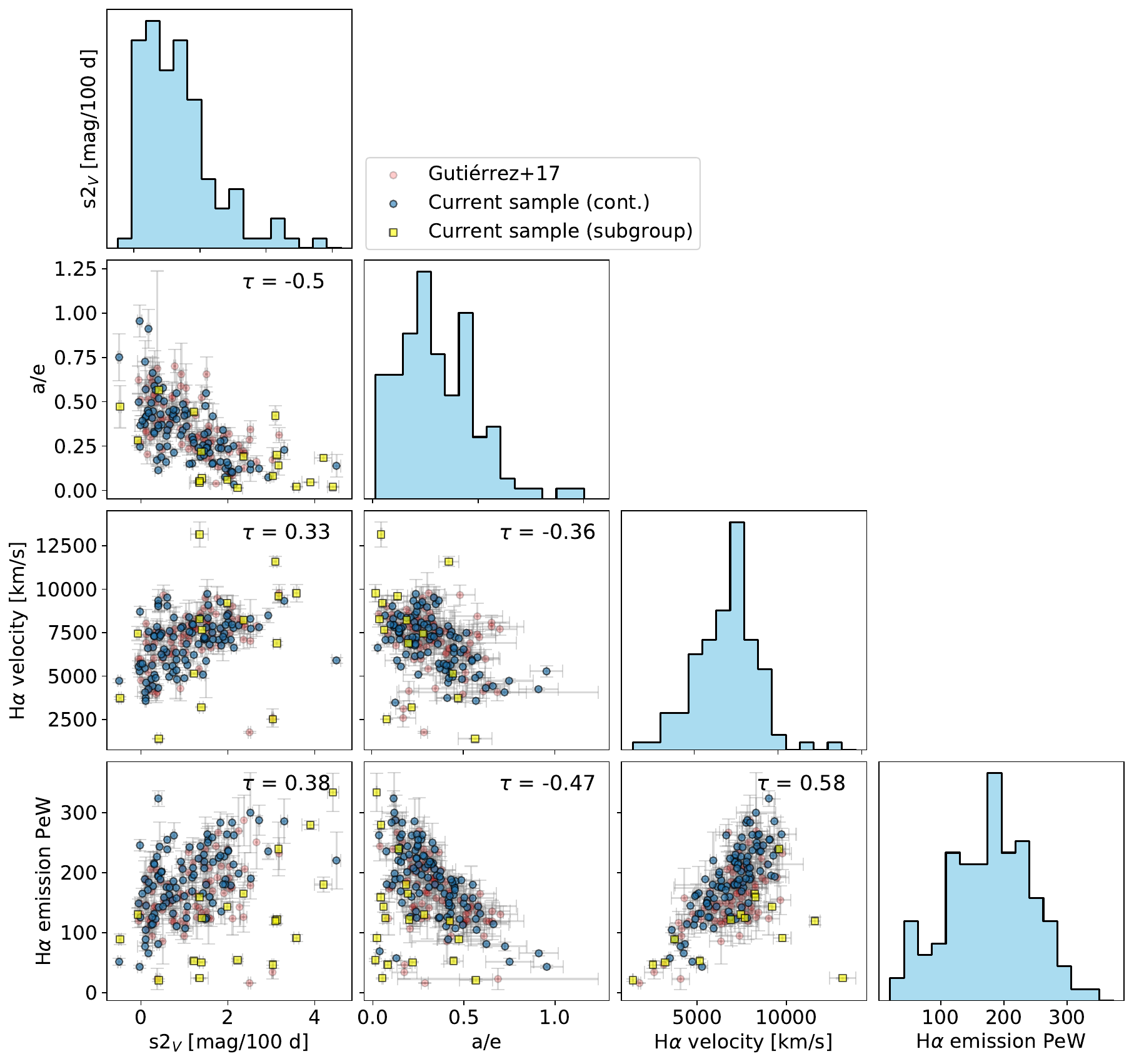}
\caption{Cornerplot of the measured spectral and light curve parameters for our supernova sample (blue and yellow squares together) at the epoch of 50 days. The blue circles represent the supernovae that belong to the 'continuous' subclass of SNe II, while the yellow squares are the objects that are members of the detached subgroup based on their spectral appearance and the clustering performed in Sec.~\ref{sec:clustering}. The red circles represent the measurements done in \cite{Gutierrez2017a}, for comparison. The $\tau$ values are the Kendall-tau correlation parameters for the canonical SNe.}
\label{fig:phys_params}
\end{figure*}

To compare the spectral properties to a number of easily measurable physical parameters, we estimated the light curve decline rate in the photospheric phase ($s_2$), the \ha\ absorption-to-emission ratio ($a/e$), and the \ha\ absorption line velocity ($v_{\textrm{H}\alpha}$). For the $s_2$ parameter, we adopted published $V$-band decline rates wherever available, most notably from \cite{Anderson2014a} for the CSP-I sample. These values and their sources are listed in Table~\ref{tab:sample1}. For SNe lacking literature values, we measured $s_2$ directly following the methodology outlined in \cite{Anderson2014a}, ensuring consistency across the dataset. Fig.~\ref{fig:lightcurves} in the Appendix shows the sample light curves, with the subgroups defined in Sec.~\ref{sec:clustering} highlighted for comparison.
For the spectroscopic parameters, we developed an automated and uniform fitting method applied to all sample spectra. To measure line velocities, we first estimated the local continuum slope around the \ha\ absorption feature, using the continuum fitting approach described in Section~\ref{sec:contrem}. This slope served as the basis for estimating measurement uncertainties: we generated 100 realizations of each spectrum by applying small random tilts, reflecting the continuum variability. For each realization, we fitted the \ha\ absorption profile using a Generalized Additive Model (GAM)-based approach \citep{Hastie1986}, which effectively followed the change in the line profile. GAMs, constructed from linear combinations of spline functions, are non-parametric and flexible. They are well-suited to asymmetric or irregular line profiles without requiring specific profile assumptions (e.g., Gaussian or Lorentzian). The fits were performed over the absorption component of the \ha\ P-Cygni profile, between 6250 \AA\ and the maximum of the line, which was set iteratively. We defined the absorption line velocity as the wavelength corresponding to the minimum of the fitted profile. While this measurement could have been conducted on the continuum-normalized spectra, we retained the original continuum to ensure consistency with the methodology of \cite{Gutierrez2017a}. 

To determine \ha\ ($a/e$), we first estimated the local continuum across the line profile by fitting a linear baseline to the edges of the feature, identified using the continuum-normalized spectra. We opted not to measure the $a/e$ directly from the normalized spectra, as the continuum is often curved underneath the line, which could potentially introduce bias. Once the continuum was defined, we integrated the absorption and emission components to compute their pseudo-equivalent widths (pEWs) using the \texttt{simps} integration method from \textsc{scipy} package \citep{Virtanen20}.

Fig.~\ref{fig:phys_params} shows a corner plot comparing the derived parameters for our SN sample. For the CSP-I sample, we also display literature values from \cite{Gutierrez2014} and \cite{Gutierrez2017a} for validation. Overall, we find good agreement between our measurements and those from the literature. A small systematic offset of $\sim 10\%$ in the \ha\ emission strength relative to \cite{Gutierrez2017a} is observed, likely due to differences in the continuum placement during the pEW estimation. Our re-analysis further confirms known correlations among the parameters, most notably between $s_2$ and $a/e$, reinforcing the link between light curve decline rates and spectroscopic evolution. Importantly, these observables show a clear continuity rather than a dichotomy, supporting that SNe~IIP and IIL represent opposite edges of the same continuous SN class. This re-measurement is both a partial confirmation and an important sanity check for the results of our data-driven approach.

\subsection{Spectral differences for IIP/L: gradual change of features across light curve decline rates}
\label{sec:reconstruction}

A long-standing question in the study of SNe~II is how spectral features correlate with the light curve decline rate. Previous empirical studies have largely focused on the most prominent optical emission line, \ha, finding that for fast-declining (IIL) SNe, the \ha\ line is increasingly blueshifted and its absorption profile weakens \citep{Anderson2014b, Gutierrez2014, Gutierrez2017}. We independently confirm these effects and uncover further correlations across the full optical spectral range.

\begin{figure*}
\centering
\includegraphics[width = 0.97\linewidth]{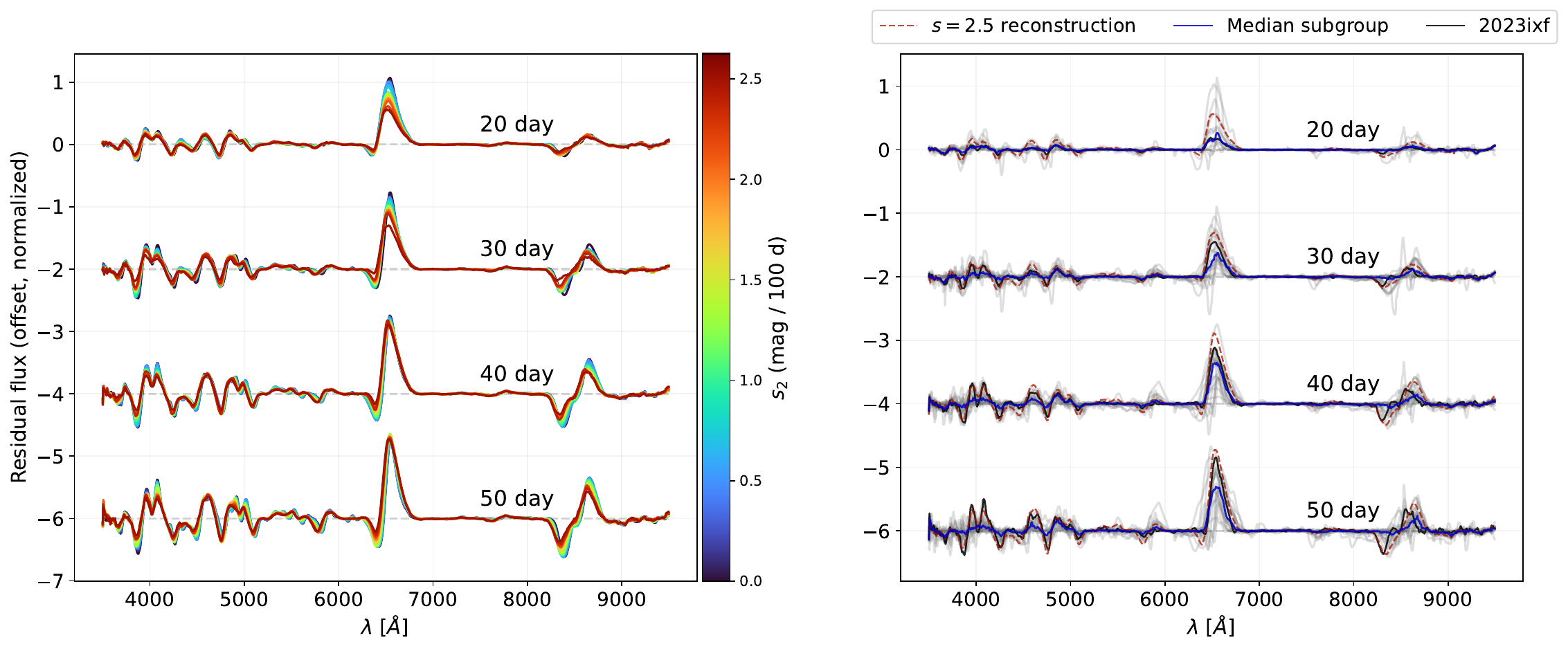}
\caption{\textbf{\geza{Top}:} Median reconstructed spectral trends as a function of the $s_2$ light curve decline rate based on our PCA analysis. The four sets of spectra correspond to the four investigated epochs. The color of the individual spectra shows the $s_2$ light curve decline rate for which they were calculated. \textbf{\geza{Bottom}:} Subgroup spectra. The gray spectra show those of the subgroup objects (defined by Fig.~\ref{fig:tsne}, the blue spectra show their median for each epoch. The red dashed spectrum shows the reconstructed spectrum for $s = 2.5$ [mag/100 d] (which is the darkest red spectrum on the left-hand side plot), while the black curve shows the spectra of SN 2023ixf.}
\label{fig:trends_w_s2}
\end{figure*}

By normalizing and interpolating the spectra, we were able to systematically and simultaneously examine how spectral features vary with the light curve $s_2$ across the entire wavelength range. This was done by computing a moving median of each PCA coefficient as a function of  $s_2$ (ranging from 0 to 2.5 [mag/100 d]), using a window width of 1 [mag/100 d]. The median spectra were then reconstructed by linearly combining the corresponding PCs. Since the individual principal components are linearly independent (by definition of PCA), this reconstruction could be performed by taking the sum. Objects that formed a detached subgroup in Fig.~\ref{fig:tsne} were excluded from this analysis.

Although this method cannot capture the full spectral diversity, only mean trends, it reveals a clear connection between $s_2$ and the spectra. As shown in Fig.~\ref{fig:trends_w_s2}, significant correlations exist even among spectroscopically "normal" SNe~II. The strength and nature of these correlations evolve with time: at early epochs (e.g. day 20), \ha\ shows a strong dependence on $s_2$, while at later epochs (e.g. at day 50), differences become more subtle. Besides \ha, the Ca II near-infrared (NIR) triplet also varies with the decline rate, and correlated changes are also observed across the metal line blanketing region. Notably, the strengths of H$\gamma$ and the Fe II lines seem to correlate with $s_2$, while H$\beta$ appears largely unaffected.

We note that establishing the connection between PCs and external parameters like $s_2$ is inherently flexible and can be approached in multiple ways. 
While PCA facilitates dimensionality reduction, outlier identification, and reconstruction, the trends present between the spectral appearance and the decline rate can be qualitatively shown without applying it. This is shown in Appendix~\ref{app:noPCA}, in Fig.~\ref{fig:IIP-IIL}, where we plot the sample spectra between epochs 25 and 35 days for decline rates below and over $s_2 = 1.5$ [mag/100 d], both before and after the continuum normalization. The simple check aligns well with the trends recovered by the PCA analysis; larger decline rates lead to more suppressed spectral features.

We further tested the robustness of the trends shown in Fig.~\ref{fig:trends_w_s2} by evaluating their dependence on the chosen median filter or regression method. In addition to the moving median approach, we explored two alternative techniques: fitting a linear function to the PC -- $s_2$ relationship, and applying a random forest regressor.
While the exact scaling of the spectral features is slightly different in each setup, all approaches yielded qualitatively consistent results: spectral features (especially \ha\ and the Ca II NIR triplet) are suppressed and blueshifted with higher $s_2$ values, alongside correlated changes in the metal line blanketing region.

To contrast this with the detached subgroup, the right panel of Fig.~\ref{fig:trends_w_s2} presents their denoised and interpolated spectra. 
The dominant difference is the overall suppression of spectral features in the subgroup objects compared to the continuous set. This is consistent with early-time CSM interaction, with ejecta not optically thick enough to form narrow lines.

\section{Discussion}
\label{sec:disc}

In the above sections, we demonstrated how spectral interpolation across epochs, combined with dimensionality reduction techniques, can be effectively used to uncover correlations between SN-specific parameters and spectral shapes. We employ principal component analysis as a basis for our work, similarly to other data-driven studies into SN spectra, such as \cite{Lu2023} or \cite{Burrow24}.

At the core of the method is the continuum normalization step, which significantly mitigates the impact of flux calibration uncertainties and unknown reddening. The approach successfully reproduces well-established correlations reported in earlier studies (e.g. \citealt{Gutierrez2014} and \citealt{Gutierrez2017}), while offering the added benefit of enabling comparison between spectra at fixed epochs across the entire optical range, rather than being limited to a small set of specific descriptive parameters.

As discussed in Sections~\ref{sec:clustering} and \ref{sec:reconstruction}, the spectra of SN IIP/L reveal both a degree of continuity and a certain level of dichotomy. Fig.~\ref{fig:tsne} highlights a meaningful separation: one subset of SNe displays well-defined correlations (especially in the appearance of \ha), while another group deviates from this trend, showing unusual or strongly suppressed line profiles. We note that this distinction is not clear-cut in earlier phases: the separation becomes apparent only at later phases (day 50) in our analysis. At earlier phases, the data exhibit larger variability, which smears the dichotomy observed at later phases. Many of these subgroup objects are explained through CSM interaction, such as for objects from \cite{Pessi23}, which supernovae generally also exhibit steeper light curve declines and suppressed spectral features. In addition to these cases, a few SNe display atypical behavior, for instance, SNe~2016bkv \citep{Hosseinzadeh17} or 2018zd \citep{Hiramatsu21} show narrow P-Cygni lines, while SN~2020cxd \citep{Yang2021} is notable for its peculiar late-time re-brightening during the end of the plateau phase. Projecting SNe exhibiting strong flash features from the  \cite{Jacobson-Galan2024} sample also supports the interpretation of enhanced CSM interaction, as these events show remarkable overlap with the subgroup found in the t-SNE space.

This is also supported by the comparison with the radiative transfer models from \cite{Hillier2019}. While the models assuming steady mass loss, which reduces the hydrogen envelope mass, are similar to SNe exhibiting steeper light curve declines, they cannot explain the full spectral variability in the sample. The models assuming CSM interaction encompass a larger range in spectral diversity, and similarities can be found with members of the subgroup, the 2023ixf-like transitional objects and the more canonical IIP-s as well. At later phases, however, the CSM interacting models show slightly differing spectra from canonical IIL-s, which may indicate that a more complete (and likely, more computationally expensive) treatment of mass loss and CSM interaction would be necessary in the models to fully explain the spectral diversity.

Our resampling analysis \geza{suggests} that the emergence of a subgroup is not an artifact of the t-SNE representation, and \geza{it shows moderate stability in our sample}.
However, we suspect the separation of the subgroup only reflects the low number of such objects in our sample.
In Fig.~\ref{fig:tsne} we showed that the subgroup is connected to enhanced CSM interaction, which should scale continuously, hence the subgroup is likely only one extreme of the underlying continuum.
Observing additional objects with similar or stronger CSM interacting signals than 2023ixf or 2024ggi will likely reveal a continuity towards these objects too.
It is important to note, however, subgroup objects identified in the embedding remain distinct in the direct correlations of spectroscopic observables, particularly those involving the strength of \ha~(Fig.~\ref{fig:phys_params}).
This supports the interpretation that enhanced CSM interaction significantly alters the Balmer line profiles, weakening the empirical correlations followed by canonical SNe~II. This conclusion is consistent with the recent findings of \cite{Dessart-JG2025}, who also demonstrated that the H$\alpha$ profile serves as a key tracer of CSM interaction at earlier phases.

Within the canonical subset of SNe, we recover continuous changes in the spectral appearance between SNe~IIP and IIL, as shown in Sec.~\ref{sec:clustering} and Sec.~\ref{sec:reconstruction}. This manifests as subtle but systematic trends across the $s_2$ decline rates, which are more pronounced at earlier phases, in line with the higher spectral diversity at these times found through PCA.
A plausible explanation for this early-time diversity is CSM interaction, which can suppress and delay line profiles, as also noted by \cite{Hillier2019}, thereby introducing greater variation and partially masking the intrinsic similarities among explosions. These differences likely become more subtle once the CSM is swept up by the SN ejecta and ceases being a dominant source of extra light. However, other physical factors may also contribute to the observed early phase diversity, including variations in explosion energy, envelope density profile and radius, Ni-mixing, non-linear temperature effects, all of which can affect the line profiles in the first weeks when the photosphere is probing the outer, still hot regions of the ejecta. On the other hand, at later phases, when the hydrogen recombination dominates, these effects are expected to even out, reducing the overall spectral diversity \citep{Dessart2011, Valenti16}.

As shown by the reconstructed spectra for the continuous sample (Fig.~\ref{fig:trends_w_s2}), the variation along $s_2$ manifests primarily as a progressive suppression of spectral features across the entire wavelength range at earlier phases. This effect is particularly pronounced at the 30-day epoch, where fast-declining SNe (IIL) display weaker \ha\ absorption and emission, reduced Ca II NIR triplet strength, and diminished metal line blanketing. These consistent patterns further support the interpretation that $s_2$ encodes meaningful information about the physical conditions of the explosion (e.g. hydrogen envelope mass) and the degree of interaction with the circumstellar medium (CSM). The findings align well with those indicating that the class of SNe II, especially SNe~IIP/L, forms a continuity, rather than a set of distinct subgroups. This was first suggested based on light curve \citep{Anderson2014a} and spectral properties \citep[e.g.,][]{Anderson2014b, Gutierrez2017}, and further supported by radiative transfer models \citep{Hillier2019}. More recent studies using early light curves and rise times \cite[e.g.,][]{Nagao2020, Hinds2025} and Ertini et al. in prep. have strengthened this unified scenario.

Although our spectral reconstructions focus on SNe with $s_2 = 2.5$ mag / 100 day, the steeper-declining events from \cite{Pessi23} show spectral trends observed that are consistent with the broader patterns observed. While we did not include these objects for reconstructing the average trends shown in Fig.~\ref{fig:trends_w_s2} due to a small sample size, they are included in the qualitative check in Appendix~\ref{app:noPCA}. These objects tend to represent intermediate cases between standard SNe~II and SNe~IIn, likely shaped by enhanced CSM interaction. Their alignment with the overall trends suggests that the varying degree of CSM interaction plays a key role in driving the observed changes in spectral features, particularly at the high end of the $s_2$ distribution.

Understanding the continuity and spectral diversity of SNe~II is crucial for their use as standard candles \citep{Poznanski2009,deJaeger2022}. The observed correlations between spectroscopic and photometric parameters are promising for calibrating absolute magnitudes. 
However, the presence of a spectral subgroup \geza{in our quality-selected sample} cautions against treating all SNe \geza{IIP/L} as a homogeneous population for standardization.
While we suspect the current \geza{gap is due to the low SN count} will be bridged by future observations and \geza{more complete sample}, the presence of a subgroup of objects showing enhanced CSM interaction and following trends different from the canonical sample (with all the subset objects being outliers for the correlations in Fig.~\ref{fig:phys_params}) highlight the need for additional standardization parameters that can account for their diversity.
This issue is especially relevant for applying the method to events like SN~2023ixf (see \citealt{Zheng25}). Although it benefits from accurate calibrator distances (e.g., \citealt{Huang2024}), SN~2023ixf lies on the interface of the spectroscopic subgroup, where correlations are altered.

A natural extension of the method is its potential use as a classification tool for new SNe. On the one hand, the simplest approach is to incorporate some of the interpolated spectra as templates to the currently existing classifiers, such as SNID \citep[Supernova Identification,][]{Blondin2007} or GELATO \citep{Harutyunyan2008}. This would extend these classifiers to better cover the more "irregular" or "CSM-interacting" members of the SN II population, which are currently underrepresented in standard template libraries. However, these template matching methods only perform the comparisons at a single, discrete epoch. This made sense in the past, given the limited sampling of observations. A more powerful approach would involve comparing the \emph{full-time} evolution of a supernova’s spectrum against a library of interpolated spectral series. Instead of matching a single spectrum, this method would identify the object whose spectral evolution most closely mirrors that of the target, effectively finding its “spectroscopic twin.” Our interpolation technique is ideally suited for such comparisons, allowing consistent spectral matching across all available epochs.

The idea of using spectral interpolation for classification purposes has recently been explored by other works as well: for instance, \cite{Ramirez2024} used spectral interpolation to improve photometric classification of SNe. As they cite, their method can further be used to estimate the necessary K-correction to be applied to photometry. Likewise, \cite{Vincenzi2019} demonstrated how spectrophotometric templates can be created purely on an observational basis, by interpolating high-quality spectral time series with photometry. While being the most complete approach, their method requires high-quality observations and an assumption on the total reddening towards the SN, which poses a limitation. In contrast, our approach will complement the above by not only identifying the type but also proposing a set of objects that evolve spectroscopically similar to previously observed batches, while being less susceptible to observational effects, similar to SNID.  It also offers the possibility of estimating the explosion time from spectroscopy alone, an important capability when early photometry is missing. We plan to explore these directions further in a future work.

It is important to note that the found correlations are limited by the precision in estimating the explosion time ($t_0$) for individual SNe. While this introduces statistical uncertainty, which we did not propagate into the analysis at this stage. Limiting ourselves to objects with better-defined $t_0$ reduces our exposure to this issue. Furthermore, for most of our sample, particularly spectra from the past decade, $t_0$ is known to within a few days thanks to extensive all-sky survey coverage.

Future works extending on the basis presented here will not be exposed to this problem, owing to the multitude of already working automated wide-field surveys, such as the Zwicky Transient Facility \citep[ZTF,][]{Bellm2019, Graham19}, or the Asteroid Terrestrial impact Last Alert System \citep[ATLAS,][]{Smith2020}, the Gravitational-wave Optical Transient Observer \citep[GOTO,][]{Dyer2020} or the BlackGEM survey \citep{Bloemen2016}, along with the long awaited Vera Rubin LSST survey \citep{LSST}. Combining these wide-field survey data of the transients with the rapid spectroscopic follow-up triggered by the community will yield a wealth of data where the time of explosion can be pinned down to high precision.

\section{Summary and conclusions}
\label{sec:conc}

In this paper, we utilized time-series interpolation and dimensionality reduction techniques, such as Gaussian Process-based time series reconstruction and Principal Component Analysis to uncover correlations between spectral features and light curve decline rates. Our primary goal was to investigate potential spectral differences between SNe~IIP/L, that do not exhibit long-lasting narrow emission features characteristic of SNe~IIn. We based our analysis on publicly available SN~II\geza{P/L} data, either from published surveys or more targeted SN works.

To minimize uncertainties related to flux calibration and reddening, we made use of empirical continuum normalization in our work. We then processed the resulting "feature spectral time series" by interpolating them per wavelength bin onto a fixed time grid. These interpolated spectra were subsequently analyzed with PCA to identify correlations between the spectral features and the light curve decline rates.

The main findings of our analysis can be summarized as follows:

\begin{itemize}

\item Flattening the spectra provides a simple and effective tool for combining observations from different surveys and instruments.

\item Spectral diversity of SNe IIP/L decreases with time: by 40-50 days post-explosion, SN spectra become more homogeneous (they exhibit smaller variability) compared to earlier epochs. CSM interaction is expected to be stronger at early phases and may play a major role in driving this effect; however, variations in the initial conditions of the explosion, such as its density profile, Ni-mixing and explosion energy, are also likely to contribute.

\item Most SNe~II in our sample form a continuous spectroscopic group across all features, along with a subgroup likely explained by enhanced CSM interaction. 
This interaction is strong enough to affect early spectra, but the CSM is not optically thick enough to produce long-lived narrow emission lines, \geza{as confirmed through} a comparison to CSM interacting objects from \cite{Jacobson-Galan2024}. 
\geza{This subgroup showed moderate stability ($\sim$60\% bootstrap recovery) under our resampling tests, and could be identified in a data-driven way.}
These objects do not follow the standard empirical correlations between spectroscopic properties, \geza{indicating} that CSM interaction can significantly affect line profiles.
\geza{The subgroup members likely represent the extremes of the spectroscopic distribution within the present sample, where pronounced CSM interaction-induced changes in the spectra, combined with their relatively small numbers, lead to their apparent separation from the bulk of SNe~II. Future observations of SNe~II exhibiting intermediate levels of CSM interaction will be essential for mapping the transitions required to populate the region between the subgroup and the bulk of the sample.}

\item We observe robust correlations for the rest of the sample between spectral features and the light curve decline rate $s_2$, a proxy for distinguishing Type IIP or IIL SNe. We find that the strength of most spectral features scales with this parameter. We recover past results, such as faster-declining SNe (IIL-like), systematically exhibiting diminished emission and sometimes even lacking absorption lines. This trend is visible not only in H$\alpha$, but also in the Ca II NIR triplet and the metal line blanketing region. Additionally, we also recover the known blueshifts in line profiles across the spectrum.

\item Spectral comparisons with nearby events (SN~2023ixf and SN~2024ggi) and radiative transfer models from \cite{Dessart13,Dessart17,Hillier2019} agree with the trends observed in our sample. The comparison with radiative transfer models suggests that both mass loss and CSM interaction are necessary to explain the full range of spectral diversity.

\item Our analysis can be further extended towards a classification method, making use of the full spectral time series at once. This allows not only the identification of SN types, but also the discovery of spectroscopic "twins" from existing samples and a refined estimate of the explosion epoch based on spectral evolution.
\end{itemize}

In the future, the method will be further refined and extended for other types of SNe,  improving classification tools, and exploring spectral correlations in greater depth. These developments will be valuable for upcoming surveys, such as the Roman Space Telescope Core Community Survey, LSST follow-up efforts, and other transient discovery programs. Finally, tailoring this analysis for SNe II will allow us to further standardize these objects as standardizable candles or to find objects with little or limited CSM interaction that can be modeled reliably for use as cosmological probes.

\begin{acknowledgements}

The authors would like to thank the anonymous referee for the comments that helped improve the manuscript. GCS acknowledges the generous computational and financial support provided by the European Southern Observatory and the Max Planck Institute for Astrophysics. GCS thanks and appreciates the enlightening and extremely helpful discussions about this project with Christian Vogl, Stefan Taubenberger, and Bálint Seli. CPG acknowledges financial support from the Secretary of Universities and Research (Government of Catalonia) and by the Horizon 2020 Research and Innovation Programme of the European Union under the Marie Sk\l{}odowska-Curie and the Beatriu de Pin\'os 2021 BP 00168 programme, from the Spanish Ministerio de Ciencia e Innovaci\'on (MCIN) and the Agencia Estatal de Investigaci\'on (AEI) 10.13039/501100011033 under the PID2023-151307NB-I00 SNNEXT project, from Centro Superior de Investigaciones Cient\'ificas (CSIC) under the PIE project 20215AT016 and the program Unidad de Excelencia Mar\'ia de Maeztu CEX2020-001058-M, and from the Departament de Recerca i Universitats de la Generalitat de Catalunya through the 2021-SGR-01270 grant.

\end{acknowledgements}

\section*{Software}
\small
This research has made use of the \textsc{python} packages: \textsc{numpy} \citep{numpy}, \textsc{matplotlib} \citep{matplotlib}, \textsc{scipy} \citep{scipy}, \textsc{george} \citep{Ambikasaran2015}, \textsc{astropy} \citep{astropy} and \textsc{pandas} \citep{pandas}.

\section*{Data availability}
The data, the intermediate results, the interpolated spectral time series, and the code used for the analysis are openly available on the GitHub page of the author \url{https://github.com/csogeza/IIP-L-diversity}.

\bibliographystyle{aa}
\bibliography{citations}

\appendix

\section{Analysis on original data without denoising the spectra}
\label{app:noisyrun}

\begin{figure}
    \centering
    \includegraphics[width=\linewidth]{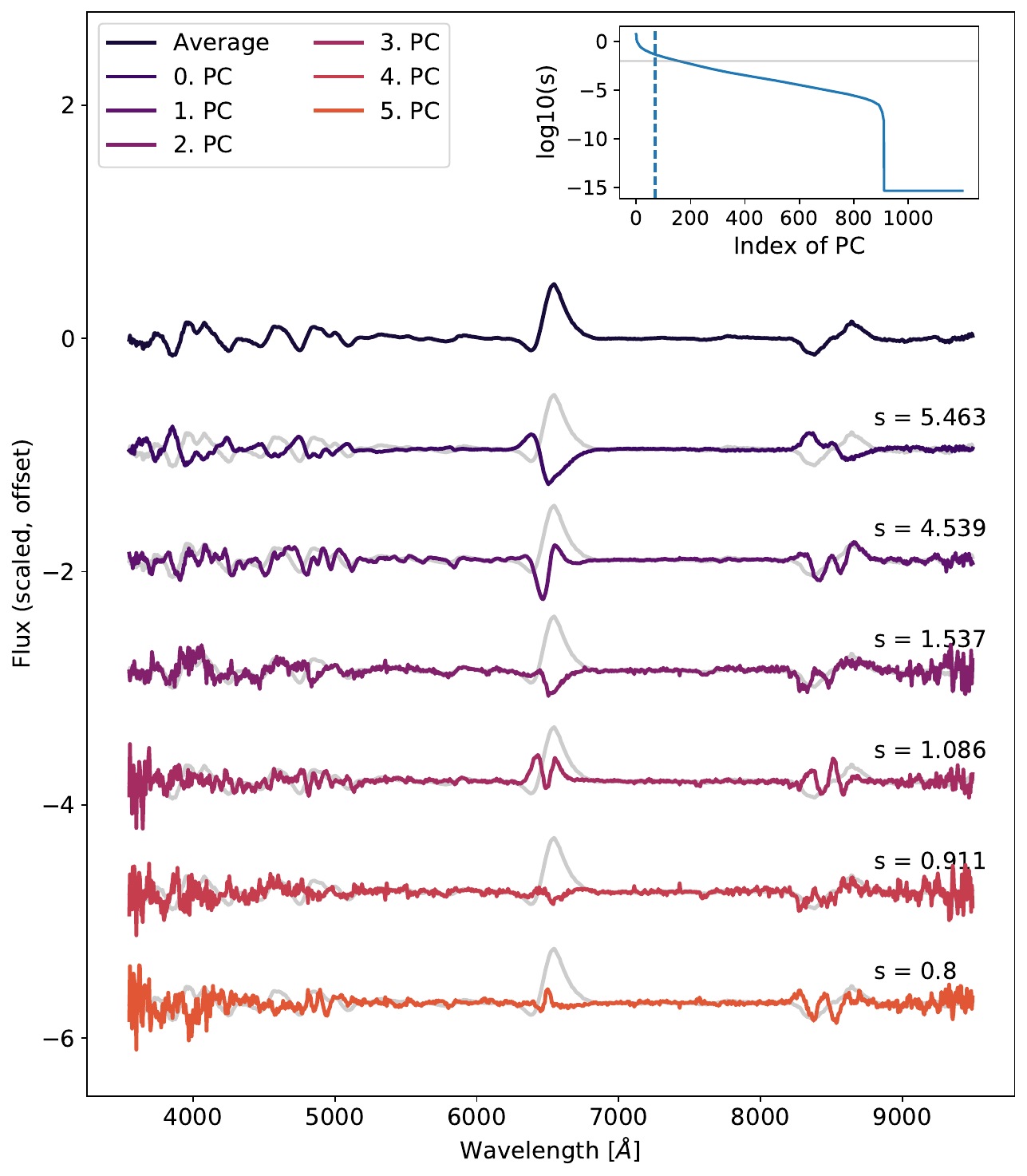}
    \caption{PCA solution for the 30-day epoch spectra without denoising them in advance. The overall trends and the individual principal components match well those obtained in the run with the applied denoising. The main difference is the lower significance of the PC-s due to the more excessive noise terms that also enter the higher-order components, meaning that more components are necessary for a more reliable decomposition.}
    \label{fig:30nodenoise}
\end{figure}

\begin{figure}
    \centering
    \includegraphics[width=\linewidth]{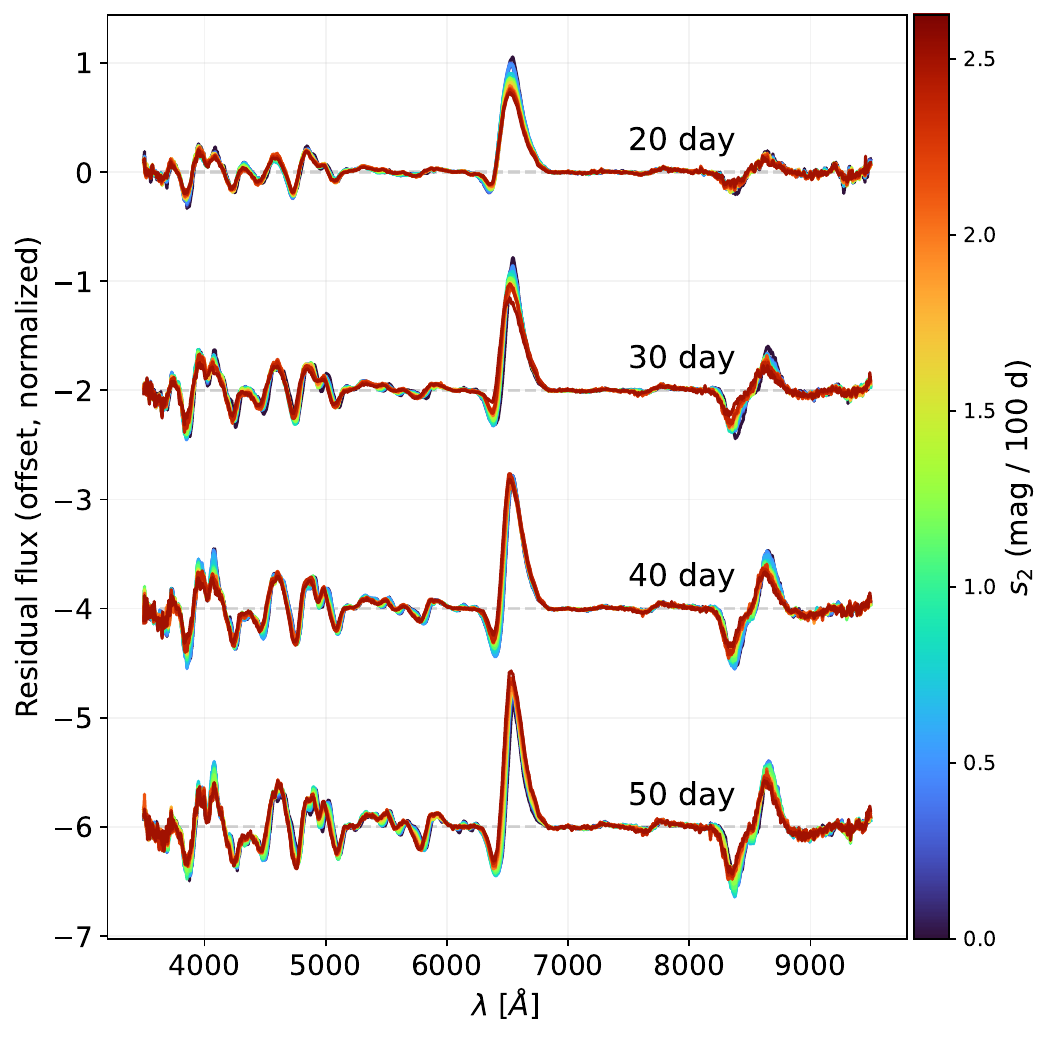}
    \caption{Recovered spectral features along changes in the light curve decline rate without the use of prior denoising on the spectra. The change in the features along the light curve decline rates matches that presented in Fig.~\ref{fig:trends_w_s2}: for more significant decline rates, one finds more suppressed emission and absorption lines.}
    \label{fig:trends-nodenoise}
\end{figure}

To provide a comparison to the PCA analysis presented in Sec.~\ref{sec:analysis}, and to show the benefits of the denoising described in Sec .~\ref {sec:denoise}, we repeat the analysis without applying the GP denoising filter on the input spectra. The outcomes of this setup are shown in Figures.~\ref{fig:30nodenoise} and ~\ref{fig:trends-nodenoise}.

The principal components displayed in Fig.~\ref{fig:30nodenoise} resemble those from the denoised case. However, the impact of noise is evident: even the higher-order components display strong noise patterns towards the red end of the spectra. In addition, the significance spectrum changes substantially; it becomes shallower compared to that of Fig.~\ref{fig:PCA_eigenvectors}, indicating that more PCs are required to explain the same amount of variance. The eigenvalue cutoff also shifts to higher PC indices in the absence of denoising.

Despite these differences, the procedure recovers the same trends between the spectra and the decline rates, as displayed in Fig.~\ref{fig:trends-nodenoise}. Spectra associated with higher s$_2$ values exhibit more diminished lines, just as shown before. However, the recovered correlations between the change in the spectral profiles and the light curve decline rate are slightly weaker than with denoising. This demonstrates that denoising indeed does not alter intrinsic correlations that can be extracted from the data, but rather strengthens them by removing the correlated effects in the noise (e.g., fringing at the redder wavelengths), leading to a more efficient PCA decomposition. 

Finally, we note that our dataset consists of nearby SNe with generally high-quality spectra and overall not so dominant noise components. In contrast, future surveys such as those conducted with the Roman Space Telescope will observe more distant SNe with less optimized follow-up. In those cases, denoising techniques will become even more critical for maximizing the scientific return from lower signal-to-noise spectral data.

\section{Spectral differences between IIP/L without the application of PCA}
\label{app:noPCA}

\begin{figure*}
    \centering
    \includegraphics[width=0.8\linewidth]{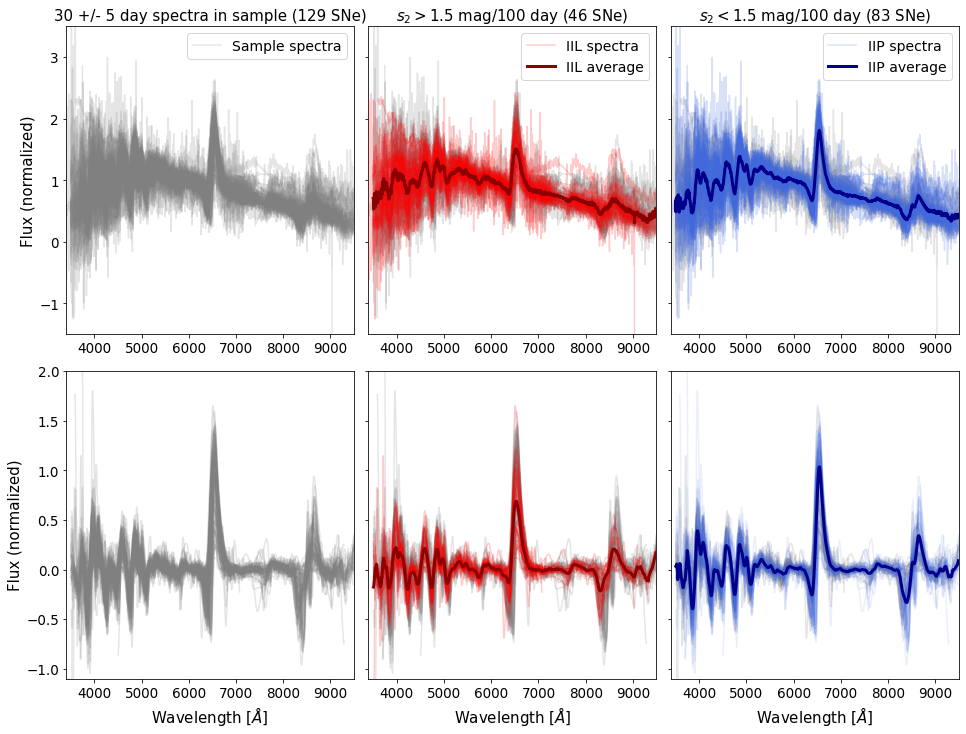}
    \caption{Spectral differences of IIP and IIL supernovae. The rows correspond to the spectra before and after the continuum normalization. The columns, from left to right: the full sample, the subsample of IIL type supernovae (exhibiting light curve decline rates larger than 1.5 mag / 100 d), and the subsample of IIP supernovae with lower decline rates. The bold coloured lines show the medians of the subsamples.}
    \label{fig:IIP-IIL}
\end{figure*}

Ultimately, it is not necessary to apply interpolation or PCA decomposition to qualitatively assess the relationship between spectral features and light curve decline rates. To demonstrate that the recovered trends are not a result of these methods, we divided the SNe into two groups based on their light curve decline rates, using a threshold of 1.5 mag / 100 day to separate the more canonical IIP-s from IIL-s). Fig.~\ref{fig:IIP-IIL} shows the averaged spectra for these two groups at an epoch of 30 +/- 5 days. The comparison reveals that, on average, SNe~IIL exhibit more diminished features of hydrogen, along with subtle suppression in the metal line blanketing region and for the Ca II NIR triplet. This low-level test confirms that the spectral differences between SNe~IIP and IIL are present directly in the raw data as well, and are not the result of the applied procedure.

\section{Principal component analysis results}
\label{app:figures}

\begin{figure}
\centering
\includegraphics[width = 0.9\linewidth]{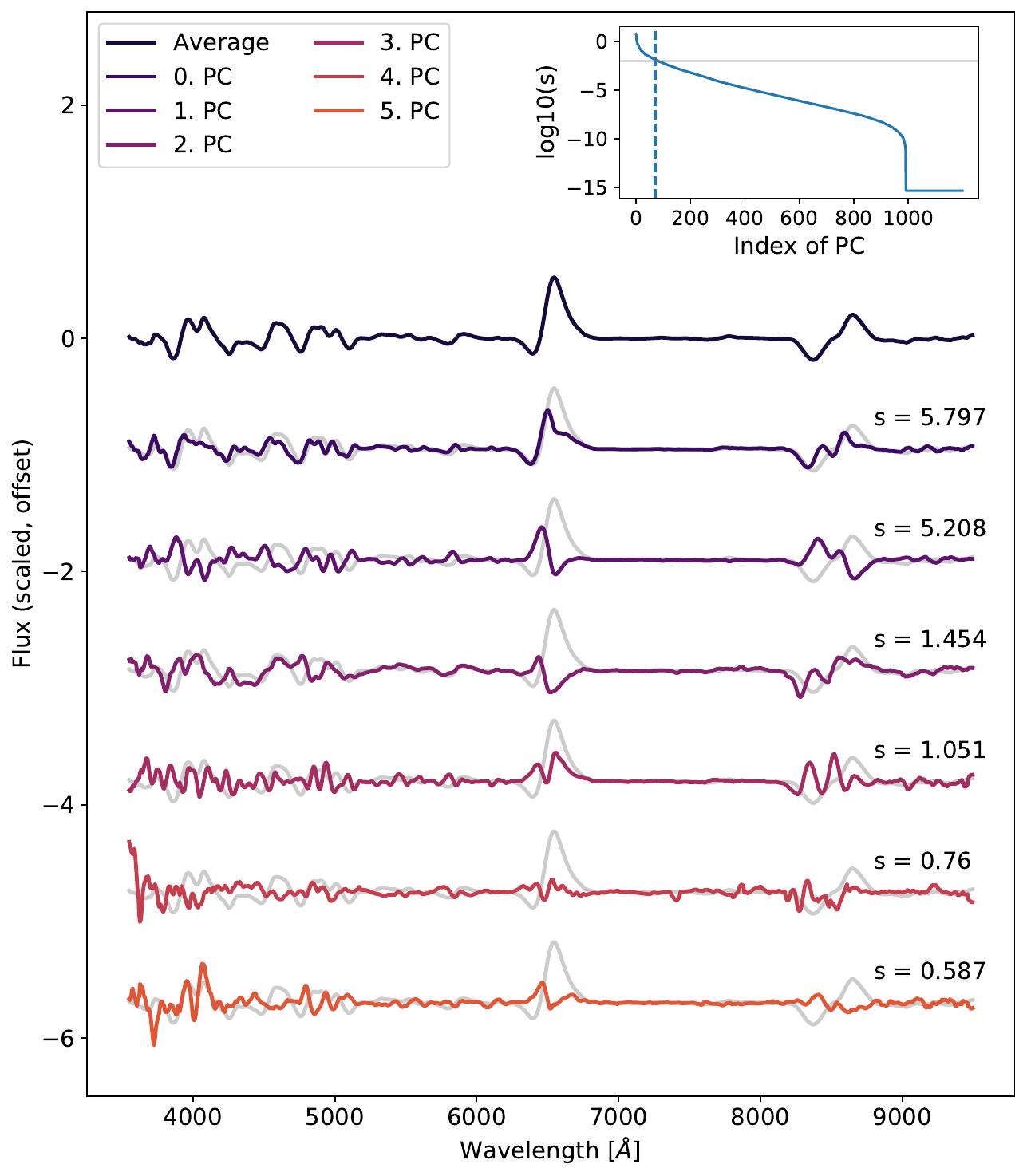}
\caption{Same as Figure~\ref{fig:PCA_eigenvectors}, but at 40 days post-explosion.}
\label{fig:PCA_eigenvectors40d}
\end{figure}

Details on the principal component analysis applied to the spectra. Table~\ref{tab:PCA} displays the singular values estimated by PCA, which track the significance of each eigenvector. As an extension of Fig.~\ref{fig:PCA_eigenvectors}, Fig.~\ref{fig:PCA_eigenvectors40d} presents the PCA eigenvectors obtained for the spectra at the 40-day epoch.

\begin{table}
\centering
\small
\begin{tabular}{c||cccc}
& \multicolumn{4}{c}{\textbf{Epoch}}\\
\cline{2-5}
\textbf{PC order} & 20 d & 30 d & 40 d & 50 d\\
\hline
\hline
1 & 4.59 & 5.11 & 5.80 & 6.77 \\
2 & 2.84 & 4.17 & 5.21 & 6.53 \\
3 & 1.09 & 1.18 & 1.45 & 1.70 \\
4 & 0.54 & 0.81 & 1.05 & 1.20 \\
5 & 0.40 & 0.47 & 0.76 & 1.01 \\
6 & 0.36 & 0.41 & 0.59 & 0.78 \\
\hline
\end{tabular}
\vspace{0.2cm}
\caption{Singular values estimated by the PCA for individual epochs. While the singular values estimated for the various epochs are not directly comparable, they show how the ratios and weights of the individual eigenspectra change during the evolution.}
\label{tab:PCA}
\end{table}

\section{Sample light curves}
\label{app:LCs}

\begin{figure}
\centering
\includegraphics[width = 0.9\linewidth]{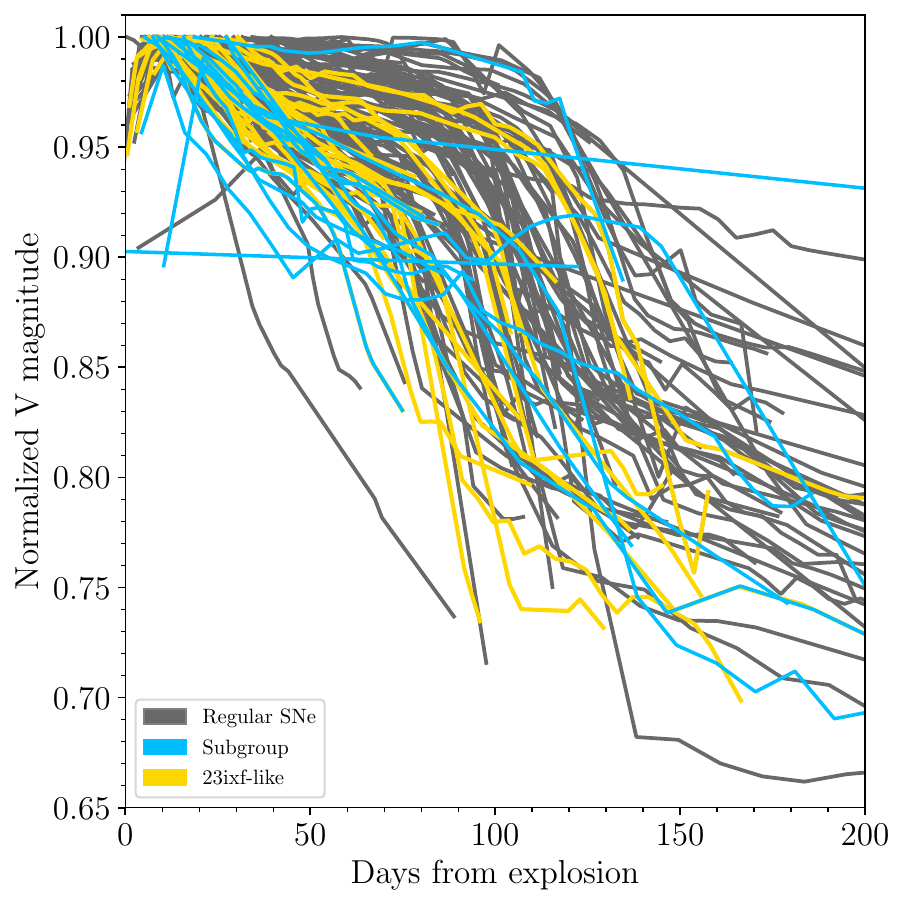}
\caption{V-band light curves of our sample objects. The gray curves correspond to the "continuous" sample, the blue curves to the subgroup objects, and the yellow curves to the objects that are similar to 2023ixf in their spectral appearance. The light curves have been normalized to their peak/maximum observed brightness to better visualize differences.}
\label{fig:lightcurves}
\end{figure}

Fig.~\ref{fig:lightcurves} shows the normalized sample light curves color-coded according to the groups they are members of as defined by the t-SNE projection. The plot clearly illustrates that most of the light curves with long plateaus correspond to SNe belonging to the continuous or "normal" group, while members of the subgroup predominantly exhibit faster decline rates. Combined with their spectral signatures, these steeper declines further support the interpretation that enhanced CSM interaction is the key driver of the detached group. One notable exception is SN~2003E, which shows a long plateau; it was marked as "irregular" due to its unusual \ha\ behavior, where the line temporarily weakens during the photospheric phase \citep{Gutierrez2017a}. We have also highlighted a subset of objects with spectra similar to SN~2023ixf, which show intermediate decline rates and correspond to the more canonical SNe~IIL. The variability in the decline rate within this spectral group may reflect different degrees of moderate CSM interaction (potentially driven by variations in CSM density or extent) or intrinsic differences in the ejecta (mass of the hydrogen envelope, mass loss). However, since most of these 2023ixf-like objects display relatively steep declines, and given that SN~2023ixf itself was influenced by CSM interaction \citep{Zheng25}, we consider CSM interaction to be a likely significant contributor within this subset as well.

\newpage

\section{Sample summary}
\label{ap:tables}

Tables~\ref{tab:sample1}, \ref{tab:sample2}, and \ref{tab:sample3} list the sample objects and the properties used or measured for the analysis. The codes for the references are listed in the caption of the last segment, Table~\ref{tab:sample3}. Finally, Table~\ref{tab:PCA-results} lists the principal component coefficient values per supernova, along with indicating which SN belongs to the detected subset. The full tables are also available on the GitHub page of the author.

\begin{table*}
\centering
\begin{tabular}{lccccccc}
\toprule
SN & Outlier & Epoch [d] & PC-1 & PC-2 & \hspace{2cm}...\hspace{2cm} & PC-69 & PC-70 \\
\midrule
1986L & no & 20 & 1.798 & -0.089 & \hspace{2cm}...\hspace{2cm} & 0.069 & 0.065 \\
 &  & 30 & 0.947 & -0.597 & \hspace{2cm}...\hspace{2cm} & -0.016 & -0.018 \\
 &  & 40 & 0.191 & 1.037 & \hspace{2cm}...\hspace{2cm} & -0.082 & 0.107 \\
 &  & 50 & 0.055 & -1.247 & \hspace{2cm}...\hspace{2cm} & -0.057 & -0.013 \\
\midrule
1987A & no & 20 & -2.480 & -4.022 & \hspace{2cm}...\hspace{2cm} & -0.019 & -0.030 \\
 &  & 30 & -1.060 & 3.059 & \hspace{2cm}...\hspace{2cm} & 0.016 & 0.010 \\
 &  & 40 & -0.692 & -3.232 & \hspace{2cm}...\hspace{2cm} & 0.016 & 0.054 \\
 &  & 50 & -1.620 & 3.478 & \hspace{2cm}...\hspace{2cm} & -0.129 & -0.011 \\
\midrule
1990E & no & 20 & -0.275 & 1.561 & \hspace{2cm}...\hspace{2cm} & -0.013 & 0.004 \\
 &  & 30 & 0.216 & -1.504 & \hspace{2cm}...\hspace{2cm} & 0.051 & 0.000 \\
 &  & 40 & -0.322 & 0.718 & \hspace{2cm}...\hspace{2cm} & 0.037 & -0.027 \\
 &  & 50 & 0.774 & -0.258 & \hspace{2cm}...\hspace{2cm} & -0.056 & 0.163 \\
\multicolumn{8}{c}{\ldots} \\
\bottomrule
\end{tabular}
\caption{Excerpt table of the PCA coefficient values obtained in Sec.~\ref{sec:analysis}. The 'Outlier' column lists if the SN falls in the group of outliers defined in Fig.~\ref{fig:tsne}, while the PC-x columns list the principal component coefficient values for each of the epochs. The full table is available online, or at the GitHub page of the author \url{https://github.com/csogeza/IIP-L-diversity}.}
\label{tab:PCA-results}
\end{table*}

\newpage

\begin{table*}[ht]
\centering
\begin{tabular}{lcccccccccc}
\hline
SN & Host & z & $t_{0}$ [MJD] & \#Spec & $s_2$ [mag / 100 d] & 20 d & 30 d & 40 d & 50 d & References \\
\hline
1986L & NGC 1559 & 0.0043 & 46708.0 & 19 & 1.28(0.03) & \checkmark & \checkmark & \checkmark & \checkmark & [1] \\
1987A & LMC &   & 46850.0 & 8 &   & \checkmark & \checkmark & \checkmark & \checkmark & [2] \\
1990E & NGC 1035 & 0.00429 & 47935.1 & 5 &   & \checkmark & \checkmark & \checkmark & \checkmark & [1] \\
1990K & NGC 150 & 0.005284 & 48001.5 & 7 & 2.13(0.19) &   &   &   & \checkmark & [1] \\
1991al & NGC 4411B & 0.01525 & 48442.5 & 6 & 1.55(0.06) &   &   & \checkmark & \checkmark & [1] \\
1992af & ESO 340-G038 & 0.01847 & 48798.8 & 3 & 0.37(0.09) & \checkmark & \checkmark & \checkmark & \checkmark & [1] \\
1992ba & NGC 2082 & 0.00395 & 48884.9 & 4 & 0.73(0.02) & \checkmark & \checkmark & \checkmark & \checkmark & [1] \\
1993S & 2MASX J22522390 & 0.03301 & 49130.8 & 3 & 2.52(0.05) &   &   & \checkmark & \checkmark & [1] \\
1996al & NGC 7689 & 0.006571 & 50260.0 & 16 & 4.65(0.5) &   & \checkmark & \checkmark & \checkmark & [3] \\
1999br & NGC 4900 & 0.003201 & 51276.7 & 7 & 0.14(0.02) & \checkmark & \checkmark & \checkmark &   & [1] \\
1999cr & ESO 576-G034 & 0.02023 & 51246.5 & 5 & 0.58(0.06) & \checkmark & \checkmark & \checkmark & \checkmark & [1] \\
1999em & NGC 1637 & 0.002392 & 51474.2 & 20 & 0.31(0.02) & \checkmark & \checkmark & \checkmark & \checkmark & [4,5] \\
1999gi & NGC 3184 & 0.001961 & 51519.0 & 6 & 0.41(0.04) & \checkmark & \checkmark & \checkmark & \checkmark & [6] \\
2000eo & MCG-2-9-3 & 0.01 & 51864.3 & 14 & 4.5(0.1) & \checkmark & \checkmark & \checkmark & \checkmark & [7] \\
2002bx & IC 2461 & 0.0075 & 52364.2 & 8 & -0.02(0.01) & \checkmark & \checkmark & \checkmark &   & [7] \\
2002ew & NEAT J205430.50 & 0.02992 & 52500.6 & 6 & 3.58(0.06) &   & \checkmark & \checkmark & \checkmark & [1] \\
2002fa & NEAT J205221.51 & 0.0584 & 52502.5 & 6 & 1.58(0.1) &   & \checkmark & \checkmark & \checkmark & [1] \\
2002gd & NGC 7537 & 0.00761 & 52552.0 & 11 & 0.11(0.05) & \checkmark & \checkmark & \checkmark & \checkmark & [1] \\
2002gw & NGC 922 & 0.010368 & 52555.8 & 7 & 0.3(0.03) & \checkmark & \checkmark & \checkmark & \checkmark & [1] \\
2002hj & NPM1G +04.0097 & 0.0236 & 52562.5 & 6 & 1.92(0.03) &   & \checkmark & \checkmark & \checkmark & [1] \\
2002hx & PGC 023727 & 0.03193 & 52582.5 & 7 & 1.54(0.04) & \checkmark & \checkmark & \checkmark & \checkmark & [1] \\
2002ig &   & 0.077 & 52570.5 & 5 & 2.73(0.11) & \checkmark & \checkmark & \checkmark & \checkmark & [1] \\
2003B & NGC 1097 & 0.00369 & 52613.5 & 5 & 0.65(0.03) &   &   & \checkmark & \checkmark & [1] \\
2003E & MCG-4-12-004 & 0.0149 & 52629.5 & 4 & -0.07(0.03) & \checkmark & \checkmark & \checkmark & \checkmark & [1] \\
2003T & UGC 4864 & 0.028 & 52654.5 & 4 & 0.82(0.02) & \checkmark & \checkmark & \checkmark & \checkmark & [1] \\
2003Z & NGC 2742 & 0.00629 & 52664.5 & 7 & 0.4(0.03) & \checkmark & \checkmark & \checkmark & \checkmark & [8,9] \\
2003bl & NGC 5374 & 0.014617 & 52696.5 & 6 & 0.24(0.04) & \checkmark & \checkmark & \checkmark & \checkmark & [1] \\
2003bn & 2MASX J10023529 & 0.01277 & 52694.5 & 6 & 0.28(0.04) & \checkmark & \checkmark & \checkmark & \checkmark & [1] \\
2003ci & UGC 6212 & 0.03034 & 52711.5 & 3 & 1.79(0.04) & \checkmark & \checkmark & \checkmark & \checkmark & [1] \\
2003cn & IC 849 & 0.01808 & 52717.5 & 4 & 1.43(0.04) & \checkmark & \checkmark & \checkmark & \checkmark & [1] \\
2003cx & NEAT J135706.53 & 0.0397 & 52725.5 & 4 & 0.76(0.03) & \checkmark & \checkmark & \checkmark & \checkmark & [1] \\
2003ef & UGC 7820 & 0.01593 & 52757.5 & 5 & 0.81(0.02) &   &   & \checkmark & \checkmark & [1] \\
2003eg & NGC 4727 & 0.02612 & 52764.5 & 4 & 2.93(0.04) &   & \checkmark & \checkmark & \checkmark & [1] \\
2003ej & UGC 7820 & 0.01697 & 52775.5 & 5 & 3.46(0.05) & \checkmark & \checkmark &   &   & [1] \\
2003fb & UGC 11522 & 0.01754 & 52772.5 & 2 & 0.48(0.06) &   & \checkmark & \checkmark & \checkmark & [1] \\
2003hd & MCG-04-05-010 & 0.0395 & 52855.9 & 6 & 1.11(0.04) & \checkmark & \checkmark & \checkmark & \checkmark & [1] \\
2003hg & NGC 7771 & 0.01427 & 52865.5 & 4 & 0.59(0.03) & \checkmark & \checkmark & \checkmark & \checkmark & [1] \\
2003hk & NGC 1085 & 0.02265 & 52866.8 & 4 & 1.85(0.06) & \checkmark & \checkmark & \checkmark & \checkmark & [1] \\
2003hl & NGC 772 & 0.00734 & 52864.6 & 5 & 0.74(0.01) & \checkmark & \checkmark & \checkmark & \checkmark & [1] \\
2003hn & NGC 1448 & 0.0039 & 52870.0 & 5 & 1.46(0.02) & \checkmark & \checkmark & \checkmark & \checkmark & [1] \\
2003ho & ESO 235-G58 & 0.01438 & 52848.5 & 4 &   &   &   &   & \checkmark & [1] \\
2003ib & MCG-04-48-15 & 0.02482 & 52891.5 & 5 & 1.66(0.05) & \checkmark & \checkmark & \checkmark & \checkmark & [1] \\
2003ip & UGC 327 & 0.01801 & 52896.5 & 4 & 2.01(0.03) & \checkmark & \checkmark & \checkmark & \checkmark & [1] \\
2003iq & NGC 772 & 0.00734 & 52919.2 & 7 & 0.75(0.03) & \checkmark & \checkmark & \checkmark & \checkmark & [1] \\
2004ej & NGC 3095 & 0.0091 & 53223.9 & 3 & 1.07(0.04) &   &   & \checkmark & \checkmark & [1] \\
2004er & MCG-01-7-24 & 0.01471 & 53271.8 & 5 & 0.4(0.03) & \checkmark & \checkmark & \checkmark & \checkmark & [1] \\
2004et & NGC 6946 & 0.00016 & 53270.7 & 19 & 0.66(0.02) & \checkmark & \checkmark & \checkmark & \checkmark & [10] \\
2004fc & NGC 701 & 0.00622 & 53293.5 & 7 & 0.82(0.02) & \checkmark & \checkmark & \checkmark & \checkmark & [1] \\
2004fx & MCG-02-14-3 & 0.0089 & 53303.5 & 6 & 0.09(0.03) &   & \checkmark & \checkmark & \checkmark & [1] \\
2005J & NGC 4012 & 0.01388 & 53379.8 & 7 & 0.96(0.02) & \checkmark & \checkmark & \checkmark & \checkmark & [1] \\
2005Z & NGC 3363 & 0.019 & 53396.7 & 8 & 1.83(0.01) & \checkmark & \checkmark & \checkmark & \checkmark & [1] \\
\hline
\end{tabular}
\caption{List of supernovae used for this study. The full list of references is given below, in the caption of Tab.~\ref{tab:sample3}.}
\label{tab:sample1}
\end{table*}

\clearpage

\begin{table*}[ht]
\centering
\begin{tabular}{lcccccccccc}
\hline
SN & Host & z & $t_{0}$ [MJD] & \#Spec & $s_2$ [mag / 100 d] & 20 d & 30 d & 40 d & 50 d & References \\
\hline
2005an & ESO 506-G11 & 0.011 & 53431.8 & 7 & 1.89(0.05) & \checkmark & \checkmark & \checkmark &   & [1] \\
2005ay & NGC 3938 & 0.0027 & 53452.5 & 6 & 0.43(0.03) & \checkmark & \checkmark & \checkmark &   & [7] \\
2005cs & M 51 & 0.002 & 53549.3 & 14 & -0.03(0.04) & \checkmark & \checkmark & \checkmark & \checkmark & [11,12] \\
2005dk & IC 4882 & 0.0157 & 53601.5 & 5 & 1.18(0.07) &   &   & \checkmark & \checkmark & [1] \\
2005dn & NGC 6861 & 0.011 & 53602.6 & 4 & 1.53(0.02) &   &   & \checkmark & \checkmark & [1] \\
2005dz & UGC 12717 & 0.019 & 53619.5 & 3 & 0.43(0.04) & \checkmark & \checkmark & \checkmark & \checkmark & [1] \\
2006Y &   & 0.0336 & 53766.1 & 8 & 1.99(0.12) & \checkmark & \checkmark & \checkmark & \checkmark & [1] \\
2006ai & ESO 005-G009 & 0.0158 & 53781.6 & 11 & 2.07(0.04) & \checkmark & \checkmark & \checkmark & \checkmark & [1] \\
2006be & IC 4582 & 0.0071 & 53802.8 & 3 & 0.67(0.02) &   & \checkmark & \checkmark &   & [1,7] \\
2006ee & NGC 774 & 0.015 & 53961.9 & 3 & 0.27(0.02) &   &   &   & \checkmark & [1] \\
2006it & NGC 6956 & 0.0155 & 54006.5 & 5 & 1.19(0.13) & \checkmark & \checkmark &   &   & [1] \\
2006iw & 2MASX J23211915 & 0.03073 & 54010.7 & 4 & 1.05(0.03) & \checkmark & \checkmark & \checkmark & \checkmark & [1] \\
2007P & ESO 566-G36 & 0.041 & 54118.7 & 7 & 2.36(0.04) & \checkmark & \checkmark & \checkmark & \checkmark & [1] \\
2007U & ESO 552-G65 & 0.026 & 54133.6 & 6 & 1.18(0.01) & \checkmark & \checkmark & \checkmark & \checkmark & [1] \\
2007W & NGC 5105 & 0.0097 & 54130.8 & 5 & 0.12(0.04) & \checkmark & \checkmark & \checkmark & \checkmark & [1] \\
2007X & ESO 385-G32 & 0.0095 & 54143.5 & 9 & 1.37(0.03) & \checkmark & \checkmark & \checkmark & \checkmark & [1] \\
2007aa & NGC 4030 & 0.004887 & 54126.7 & 11 & -0.05(0.02) &   & \checkmark & \checkmark & \checkmark & [1,7] \\
2007ab & MCG-01.43-2 & 0.024 & 54123.9 & 9 & 3.3(0.08) &   & \checkmark & \checkmark & \checkmark & [1] \\
2007av & NGC 3279 & 0.00464 & 54173.8 & 5 & 0.97(0.02) & \checkmark & \checkmark & \checkmark & \checkmark & [1,7] \\
2007bf & UGC 9121 & 0.0178 & 54191.5 & 5 & 0.06(0.13) & \checkmark & \checkmark &   &   & [1,7] \\
2007hm & SDSS J205755 & 0.02 & 54336.6 & 5 & 1.45(0.04) & \checkmark & \checkmark & \checkmark & \checkmark & [1] \\
2007il & IC 1704 & 0.021 & 54349.8 & 9 & 0.31(0.02) & \checkmark & \checkmark & \checkmark & \checkmark & [1] \\
2007ld &   & 0.025 & 54376.5 & 5 & 1.12(0.16) & \checkmark & \checkmark &   &   & [1] \\
2007oc & NGC 7418 & 0.007 & 54388.5 & 13 & 1.83(0.01) & \checkmark & \checkmark & \checkmark & \checkmark & [1] \\
2007od & UGC 12846 & 0.00586 & 54400.6 & 22 & 1.55(0.01) & \checkmark & \checkmark & \checkmark & \checkmark & [1,7] \\
2007pk & NGC 579 & 0.0167 & 54414.2 & 5 & 2.14(0.06) & \checkmark & \checkmark & \checkmark & \checkmark & [1,7] \\
2007sq & MCG-03-23-5 & 0.0153 & 54422.8 & 8 & 1.51(0.05) &   & \checkmark & \checkmark & \checkmark & [1] \\
2008K & ESO 504-G5 & 0.0267 & 54475.5 & 7 & 2.72(0.02) & \checkmark & \checkmark & \checkmark & \checkmark & [1] \\
2008M & ESO 121-26 & 0.0076 & 54471.7 & 10 & 1.14(0.02) & \checkmark & \checkmark & \checkmark & \checkmark & [1] \\
2008W & MCG-03-22-7 & 0.0191 & 54483.8 & 5 & 1.11(0.04) &   & \checkmark & \checkmark & \checkmark & [1] \\
2008ag & IC 4729 & 0.0148 & 54477.9 & 6 & 0.16(0.01) &   &   & \checkmark & \checkmark & [1] \\
2008aw & NGC 4939 & 0.0104 & 54517.8 & 9 & 2.25(0.03) &   & \checkmark & \checkmark & \checkmark & [1] \\
2008bh & NGC 2642 & 0.014 & 54543.5 & 6 & 1.2(0.04) & \checkmark & \checkmark & \checkmark & \checkmark & [1] \\
2008bj & MCG +08-22-20 & 0.019 & 54537.3 & 12 & 1.64(0.04) & \checkmark & \checkmark & \checkmark & \checkmark & [7] \\
2008bk & NGC 7793 & 0.000757 & 54542.9 & 10 & 0.11(0.02) &   & \checkmark & \checkmark & \checkmark & [1] \\
2008bn & NGC 4226 & 0.0242 & 54556.3 & 11 & 1.65(0.09) & \checkmark & \checkmark & \checkmark & \checkmark & [7] \\
2008br & IC 2522 & 0.0101 & 54555.7 & 4 & 0.45(0.02) & \checkmark & \checkmark & \checkmark &   & [1] \\
2008bu & ESO 586-G2 & 0.0221 & 54566.8 & 8 & 2.77(0.14) & \checkmark & \checkmark &   &   & [1] \\
2008bx & MCG +07-31-04 & 0.0084 & 54578.3 & 5 & 1.95(0.19) & \checkmark & \checkmark & \checkmark & \checkmark & [7] \\
2008fc & UGC 2883 & 0.01734 & 54698.0 & 6 &   & \checkmark & \checkmark & \checkmark & \checkmark & [13] \\
2008fq & NGC 6907 & 0.00797 & 54720.3 & 9 & 1.94(0.09) & \checkmark & \checkmark & \checkmark & \checkmark & [14] \\
2008gi & CGCG 415-004 & 0.0244 & 54742.7 & 5 & 3.13(0.08) & \checkmark & \checkmark & \checkmark & \checkmark & [1] \\
2008gr & IC 1579 & 0.0228 & 54769.6 & 4 & 2.01(0.01) & \checkmark & \checkmark & \checkmark & \checkmark & [1] \\
2008if & MCG-01-24-10 & 0.0115 & 54807.8 & 10 & 2.1(0.02) & \checkmark & \checkmark & \checkmark & \checkmark & [1,7] \\
2008in & NGC 4303 & 0.005224 & 54824.2 & 11 & 0.83(0.02) & \checkmark & \checkmark & \checkmark & \checkmark & [7] \\
2008ip & NGC 4846 & 0.01509 & 54832.4 & 6 & 1.41(0.07) &   & \checkmark & \checkmark & \checkmark & [7] \\
2009N & NGC 4487 & 0.003449 & 54846.8 & 6 & 0.34(0.01) &   & \checkmark & \checkmark & \checkmark & [1,7] \\
2009aj & ESO 221-G18 & 0.0096 & 54880.5 & 11 &   & \checkmark & \checkmark & \checkmark & \checkmark & [1] \\
2009ao & NGC 2939 & 0.0111 & 54890.7 & 7 & -0.01(0.12) & \checkmark & \checkmark & \checkmark & \checkmark & [1] \\
2009au & ESO 443-21 & 0.0094 & 54897.5 & 8 & 3.04(0.02) & \checkmark & \checkmark & \checkmark & \checkmark & [1] \\
2009bu & NGC 7408 & 0.012 & 54901.9 & 6 & 0.18(0.04) & \checkmark & \checkmark & \checkmark & \checkmark & [1] \\
\hline
\end{tabular}
\caption{Continuation of Table~\ref{tab:sample1}. The full list of references is given below, in the caption of Tab.~\ref{tab:sample3}.}
\label{tab:sample2}
\end{table*}

\clearpage

\begin{table*}[ht]
\centering
\begin{tabular}{lcccccccccc}
\hline
SN & Host & z & $t_{0}$ [MJD] & \#Spec & $s_2$ [mag / 100 d] & 20 d & 30 d & 40 d & 50 d & References \\
\hline
2009bz & UGC 9814 & 0.0108 & 54915.8 & 4 & 0.5(0.02) & \checkmark & \checkmark &   &   & [1] \\
2009dd & NGC 4088 & 0.0034 & 54925.0 & 11 & 1.48(0.09) & \checkmark & \checkmark & \checkmark & \checkmark & [7] \\
2009ib & NGC 1559 & 0.00435 & 55041.3 & 4 & 0.04(0.03) & \checkmark & \checkmark & \checkmark & \checkmark & [15] \\
2010aj & MCG -01-32-35 & 0.0212 & 55267.5 & 8 & 2.19(0.12) & \checkmark & \checkmark & \checkmark &   & [7] \\
2012A & NGC 3239 & 0.00199 & 55933.0 & 23 & 1.22(0.08) & \checkmark & \checkmark & \checkmark & \checkmark & [16] \\
2012aw & M 95 & 0.002595 & 56001.5 & 26 & 0.51(0.03) & \checkmark & \checkmark & \checkmark & \checkmark & [17] \\
2012ec & NGC 1084 & 0.0047 & 56144.0 & 27 & 0.53(0.06) & \checkmark & \checkmark & \checkmark & \checkmark & [18] \\
2013K & ESO 9-10 & 0.008066 & 56302.0 & 11 & 0.17(0.03) & \checkmark & \checkmark & \checkmark & \checkmark & [19] \\
2013L & ESO 216-39 & 0.016992 & 56312.0 & 5 & 2.22(0.11) & \checkmark & \checkmark & \checkmark & \checkmark & [19] \\
2013ab & NGC 5669 & 0.00532 & 56339.5 & 20 & 0.92(0.1) & \checkmark & \checkmark & \checkmark & \checkmark & [20] \\
2013am & NGC 3623 & 0.002692 & 56371.5 & 19 & 0.1(0.17) & \checkmark & \checkmark & \checkmark & \checkmark & [21] \\
2013ej & NGC 628 & 0.002192 & 56496.9 & 58 & 1.89(0.04) & \checkmark & \checkmark & \checkmark & \checkmark & [22,23] \\
2013fs & NGC 7610 & 0.011855 & 56571.1 & 17 & 1.12(0.08) & \checkmark & \checkmark & \checkmark & \checkmark & [24] \\
2014G & NGC 3448 & 0.0039 & 56669.2 & 15 & 2.53(0.06) & \checkmark & \checkmark & \checkmark & \checkmark & [25] \\
2014cx & NGC 337 & 0.005 & 56901.9 & 16 & 0.4(0.02) & \checkmark & \checkmark & \checkmark & \checkmark & [26] \\
2015an & IC 2367 & 0.00817 & 57268.0 & 10 & 1.22(0.13) &   & \checkmark & \checkmark & \checkmark & [27] \\
2015bs &   & 0.027 & 56922.0 & 4 & 0.4(0.08) & \checkmark & \checkmark & \checkmark & \checkmark & [28] \\
2015cz & NGC 582 & 0.0145 & 57291.1 & 9 & 1.02(0.18) & \checkmark & \checkmark & \checkmark & \checkmark & [26] \\
2016B & CGCG 012-116 & 0.004293 & 57381.5 & 12 & 0.47(0.24) & \checkmark & \checkmark & \checkmark & \checkmark & [29] \\
2016X & UGC 08041 & 0.0044 & 57405.9 & 11 & 1.33(0.08) & \checkmark & \checkmark & \checkmark & \checkmark & [30] \\
2016aqf & NGC 2101 & 0.004 & 57440.0 & 12 & 1.0(0.04) & \checkmark & \checkmark & \checkmark & \checkmark & [31] \\
2016bkv & NGC 3184 & 0.002 & 57467.3 & 16 & 0.41(0.04) & \checkmark & \checkmark & \checkmark & \checkmark & [32] \\
2016egz &  & 0.023 & 57588.2 & 8 & 1.35(0.23) & \checkmark & \checkmark & \checkmark & \checkmark & [33] \\
2016esw &  & 0.02831 & 57607.8 & 6 & 1.46(0.08) & \checkmark & \checkmark & \checkmark & \checkmark & [34] \\
2016gsd & NGC 1036 & 0.06 & 57648.0 & 7 & 3.09(0.04) &   & \checkmark & \checkmark & \checkmark & [35] \\
2017cfo &   & 0.042 & 57822.2 & 3 & 4.2(0.2) & \checkmark & \checkmark & \checkmark & \checkmark & [36] \\
2017eaw & NGC 6946 & 0.000133 & 57885.5 & 24 & 0.45(0.03) & \checkmark & \checkmark & \checkmark & \checkmark & [37] \\
2017gmr & NGC 988 & 0.005 & 57999.1 & 52 & 0.61(0.04) & \checkmark & \checkmark & \checkmark & \checkmark & [38] \\
2017gpp &   & nan & 57995.0 & 5 & 1.35(0.2) & \checkmark & \checkmark & \checkmark & \checkmark & [36] \\
2017hbj & ESO 84-21 & 0.017 & 58023.5 & 5 & 3.9(0.2) & \checkmark & \checkmark & \checkmark & \checkmark & [36] \\
2017hxz &   & 0.08 & 58048.0 & 6 & 5.56(0.4) &   & \checkmark & \checkmark & \checkmark & [36] \\
2018emt &   & 0.0239 & 58331.5 & 5 & 1.36(0.2) & \checkmark & \checkmark & \checkmark &   & [36] \\
2018eph &   & 0.03 & 58329.2 & 6 &   & \checkmark & \checkmark & \checkmark & \checkmark & [36] \\
2018hfm & & 0.0081 & 58395.2 & 9 & 4.42(0.13) & \checkmark & \checkmark & \checkmark & \checkmark & [36] \\
2018hna & UGC 7534 & 0.002408 & 58410.8 & 13 &   & \checkmark & \checkmark & \checkmark & \checkmark & [39] \\
2018lab & NGC 2207 & 0.009223 & 58480.4 & 6 & 0.26(0.1) & \checkmark & \checkmark & \checkmark & \checkmark & [40] \\
2018zd & NGC 2146 & 0.002979 & 58171.5 & 21 & 1.39(0.04) & \checkmark & \checkmark & \checkmark & \checkmark & [41] \\
2019va & UGC 8577 & 0.0088 & 58494.0 & 6 & -0.02(0.02) & \checkmark & \checkmark & \checkmark & \checkmark & [42] \\
2020cxd & NGC 6395 & 0.0039 & 58897.0 & 3 & -0.48(0.03) & \checkmark & \checkmark & \checkmark & \checkmark & [43] \\
2020fqv & NGC 4568 & 0.007 & 58938.9 & 7 & 0.67(0.05) & \checkmark &   &   &   & [44] \\
2020jfo & M 61 & 0.005224 & 58975.2 & 40 & 1.5(0.04) & \checkmark & \checkmark & \checkmark & \checkmark & [45,46] \\
2021aai & NGC 2268 & 0.007412 & 59224.0 & 7 & -0.5(0.03) & \checkmark & \checkmark & \checkmark & \checkmark & [43] \\
2021yja & NGC 1325 & 0.005307 & 59465.4 & 39 & 0.61(0.06) & \checkmark & \checkmark & \checkmark & \checkmark & [47] \\
ASASSN-15oz & HIPASS J1919-33 & 0.007 & 57261.0 & 14 & 1.98(0.04) & \checkmark & \checkmark & \checkmark & \checkmark & [48] \\
LSQ13CUW &   & 0.25 & 56593.9 & 9 & 3.17(0.02) &   & \checkmark & \checkmark & \checkmark & [49] \\
\hline
\end{tabular}
\caption{Continuation of Table~\ref{tab:sample1}. The full list of references: [1]: \cite{Gutierrez2017a}, [2]: \cite{Bouchet1989}, [3]: \cite{Benetti2016}, [4]: \cite{Leonard2001}, [5]: \cite{Elmhamdi2003}, [6]: \cite{Leonard2002}, [7]: \cite{Hicken2017}, [8]: \cite{Pastorello2003}, [9]: \cite{Knop2007}, [10]: \cite{Sahu2006}, [11]: \cite{Pastorello2006}, [12]: \cite{Pastorello2009}, [13]: private comm., [14]: \cite{Faran2014}, [15]: \cite{Inserra2015}, [16]: \cite{Tomasella2013}, [17]: \cite{Dall'Ora2014}, [18]: \cite{Maund2013}, [19]: \cite{Tomasella2018}, [20]: \cite{Bose2015}, [21]: \cite{Zhang2014}, [22]: \cite{Dhungana2016}, [23]: \cite{Bose2015ej}, [24]: \cite{Yaron2017}, [25]: \cite{Terreran2016}, [26]: \cite{Dastidar2021}, [27]: \cite{Dastidar2019an}, [28]: \cite{Anderson2018},
[29]: \cite{Dastidar2019B}, [30]: \cite{Huang2018}, [31]: \cite{Muller-Bravo2020}, [32]: \cite{Hosseinzadeh2018}, [33]: \cite{Hiramatsu2021}, [34]: \cite{deJaeger2018}, [35]: \cite{Reynolds2020}, [36]: \cite{Pessi23}, [37]: \cite{Szalai2019}, [38]: \cite{Andrews2019}, [39]: \cite{Xiang2023}, [40]: \cite{Pearson2023}, [41]: \cite{Zhang2018}, [42]: \cite{Zhang2022}, [43]: \cite{Valerin2022}, [44]: \cite{Tinyanont2022}, [45]: \cite{Sollerman2021}, [46]: \cite{Teja2022}, [47]: \cite{Vasylyev2022}, [48]: \cite{Bokstroem2019}, [49]: \cite{Gall2015}}
\label{tab:sample3}
\end{table*}


\end{document}